\documentclass[12pt]{article}

\usepackage{epsfig}
\usepackage{graphics}

\usepackage{cite,url}
\usepackage{amsmath,amssymb}

\usepackage{a4wide}
\numberwithin{equation}{section}
\DeclareMathOperator{\tr}{Tr}

\def\ket#1{\mathinner{|{#1}\rangle}}
\def\braket#1{\mathinner{\langle{#1}\rangle}}

\begin{document}
\thispagestyle{empty}
\allowdisplaybreaks
\arraycolsep0.05cm

\begin{flushright}
{\small
PITHA~09/16\\
IPPP/09/48\\
DCPT/09/96\\
SFB/CPP-09-59\\
0907.1443 [hep-ph]\\
November 13, 2009}
\end{flushright}

\vspace{\baselineskip}

\begin{center}
\vspace{0.5\baselineskip} 
\textbf{\Large\boldmath 
Soft radiation in  
heavy-particle  pair production:\\[0.02cm]
all-order colour structure and two-loop \\[0.2cm]
 anomalous dimension
}\\
\vspace{3\baselineskip}
{\sc M.~Beneke$^a$, P.~Falgari$^b$, C.~Schwinn$^b$}\\
\vspace{0.7cm}
{\sl ${}^a$Institut f\"ur Theoretische Physik E, RWTH Aachen University,\\
D--52056 Aachen, Germany\\
\vspace{0.3cm}
${}^b$IPPP, Department of Physics, University of Durham, \\
Durham DH1 3LE, England}

\vspace*{1.2cm}
\textbf{Abstract}\\ 

\vspace{1\baselineskip}
\parbox{0.9\textwidth}{ We consider the total production cross section
  of heavy coloured particle pairs in hadronic collisions at the
  production threshold.  We construct a basis in colour space that
  diagonalizes to all orders in perturbation theory the soft function, 
  which appears in a new factorization
  formula for the combined resummation of soft gluon and Coulomb gluon
  effects. This extends recent results on the structure of
  soft anomalous dimensions and allows us to determine an analytic
  expression for the two-loop soft anomalous dimension at threshold
  for all production processes of interest. 
}

\end{center}

\newpage
\setcounter{page}{1}

%%%%%%%%%%%%%%%%%%%%%%%%%%%%%%%%%%%%%%%%%%%%%%%%%%%%%%%%%%%%%%%%

\section{Introduction}
\label{sec:introduction}

Perturbative calculations of partonic cross sections at hadron
colliders often fail near the boundaries of partonic phase space due to 
logarithmically enhanced terms from soft gluon radiation. If it can be 
argued that the hadronic cross section is dominated numerically by 
these threshold logarithms, they should be summed to all orders 
in perturbation theory.  This can be done in Mellin moment 
space~\cite{Sterman:1986aj,Catani:1989ne,Kidonakis:1997gm,Kidonakis:1998nf,Bonciani:2003nt} 
or directly in momentum 
space~\cite{Becher:2006nr,Becher:2006mr,Becher:2007ty}. In either case, 
the theoretical basis for resummation is a factorization of the 
partonic hard-scattering cross section $\hat \sigma$ in the partonic 
threshold region into hard and soft contributions of the schematic form
\begin{equation}
\label{eq:factor-thresh}
\hat \sigma =H\otimes S
\end{equation}
with a hard function $H$ and a soft function $S$. The soft function 
satisfies an evolution equation, whose driving term is the anomalous 
dimension matrix of $S$. Resummation amounts to solving this 
equation.

Particularly important for experiments at LHC and Tevatron are
pair production processes of heavy 
coloured particles $H, H'$ in a collision of hadrons $N_1$ and $N_2$,
\begin{equation}
\label{eq:heavy-pair}
N_1(P_1) N_2(P_2) \rightarrow H(p_1) H'(p_2) + X.
\end{equation}
In this case the partonic cross section contains terms of the form
$[\alpha_s^n \ln^m\beta\,]$ (``threshold logarithms'') and
$(\alpha_s/\beta)^n$ (``Coulomb singularity''), where
$\beta=(1-4M^2/\hat s)^{1/2}$ is the heavy particle velocity, which
are enhanced near the partonic threshold $\hat s\approx 4 M^2$, with
$M$ the average heavy-particle mass.  The threshold logarithms have
been discussed in the past for various production processes of the
form~\eqref{eq:heavy-pair} leading to improved predictions for the
top-quark production cross section at hadron
colliders~\cite{Catani:1996yz,Bonciani:1998vc,Moch:2008qy,Cacciari:2008zb,Kidonakis:2008mu,Czakon:2008cx,Langenfeld:2009wd,Hagiwara:2008df,Kiyo:2008bv},
production of supersymmetric coloured
particles~\cite{Kulesza:2008jb,Kulesza:2009kq,Langenfeld:2009eg}, and
colour-octet scalars~\cite{Idilbi:2009cc}. In two-to-two scattering
the soft function is a matrix in colour space. In general this matrix
depends on the kinematical invariants of the scattering
process~\cite{Kidonakis:1997gm}; however, for resummation of threshold
logarithms in the total cross section, the relevant quantity is the
soft anomalous dimension at threshold. In this case, the question
arises whether the resummation of logarithms is altered by the
presence of Coulomb corrections, which must also be summed.  This has
not been addressed in the past, where the Coulomb correction is
technically considered as part of the hard function~$H$.

In a separate paper~\cite{BFS} we employ effective field theory and field 
redefinitions to derive an extended factorization formula of the form 
$\hat \sigma =H\otimes J\otimes S$ for the hard-scattering total  
cross sections related to (\ref{eq:heavy-pair}), which implies a proof
of factorization of soft gluons in the presence of Coulomb exchange. More
precisely, the cross sections of the partonic subprocesses
$p(k_1) p'(k_2)\rightarrow H(p_1)H'(p_2)+X$
where $pp'\in\{qq, q\bar q, gg, gq,g\bar q\}$ are expressed as 
\begin{equation}
\label{eq:fact}
  \hat\sigma_{pp'}(\hat s,\mu)
= \sum_{i,i'}H_{ii'}(M,\mu)
\;\int d \omega\;
\sum_{R_\alpha}\,J_{R_\alpha}(E-\frac{\omega}{2})\,
W^{R_\alpha}_{ii'}(\omega,\mu).
\end{equation}
Here $\hat{s}$ is the partonic centre-of-mass energy, and $E=\sqrt{\hat s}-2
M$.  The formula applies to heavy particle pairs produced in an S-wave and is
valid at the leading order in the non-relativistic expansion. The new function
$J_{R_\alpha}$ sums Coulomb gluon exchange related to the
attractive or repulsive Coulomb force in the irreducible colour
representations $R_\alpha$ that appear in the product representation $R\otimes
R^\prime$ of the final state particles and includes the leading Coulomb
singularities $(\alpha_s/\beta)^n$. We note that Coulomb summation has been
included in various forms in~\cite{Hagiwara:2008df,Kiyo:2008bv,Kulesza:2009kq}
where, however, the factorization of Coulomb from soft gluons is put in as an
assumption.  A formula equivalent to eq.~\eqref{eq:fact} for the factorization
of electromagnetic effects in $W$-pair production at $e^-e^+$ colliders has
been derived in~\cite{Falgari:2009zz}.  The purpose of the present paper is to
discuss the colour decomposition of the generalized soft function
$W^{R_\alpha}_{ii'}$, which is the crucial ingredient for resummation, and to
provide the two-loop anomalous dimensions and one-loop soft functions, which
are necessary for next-to-next-to-leading logarithmic (NNLL) resummations. We
shall also briefly outline the general structure of resummation and explain
further NNLL effects  not included in eq.~(\ref{eq:fact}).

The soft functions are vacuum expectation values of Wilson line 
operators that retain only information about colour and direction 
of massive and light-like particles in the hard process. The 
light-like ($n^2=0$) Wilson line for an incoming particle in the 
representation $r$ of SU(3) with generator ${\bf T}^{(r)a}$ is  
\begin{equation}
\label{eq:def-wilson-coll}
S^{(r)}_{n}(x) = \mbox{P} \exp 
\left[i g_s \int_{-\infty}^0 
d s \,n \cdot  A^a(x+n s)\,{\bf T}^{(r)a}\right].
\end{equation}
For an outgoing heavy particle in representation $R$ we define the 
time-like ($v^2=1$) Wilson line 
\begin{equation}
\label{eq:def-wilson}
S^{(R)\dagger}_{v}(x) =
\mbox{P}
\exp \left[i g_s \int_{0}^\infty d s\; v \cdot A^a(x+v s)
{\bf T}^{(R)a}\right].
\end{equation}
The inverse (adjoint) Wilson line operators follow from replacing 
path-ordering by anti path-ordering, and $i g_s\to -i g_s$. 
The soft function $W^{R_\alpha}_{ii'}(\omega,\mu)$ in 
(\ref{eq:fact}) is a descendant
of a more general soft function defined by
\begin{equation}
\label{eq:soft-general}
\hat W^{\{k\}}_{\{ab\}}(z,\mu)
=\langle 0|\overline{\mbox{T}}[ 
S_{v,b_4 k_2} S_{v,b_3 k_1} S^\dagger_{\bar{n},jb_2} S^\dagger_{n, ib_1}](z)
\mbox{T}[S_{n,a_1i} S_{\bar{n},a_2j} S^\dagger_{v,k_3 a_3} 
S^\dagger_{v,k_4 a_4}](0)|0\rangle,
\end{equation}
where $\mbox{T}$ and $\overline{\mbox{T}}$ denote time-ordering and anti
time-ordering, respectively. $n$ and $\bar n$ denote two light-like
vectors satisfying $n\cdot \bar n=2$. The superscript on the Wilson lines 
denoting the colour representation has been omitted.
In the factorization formula~\eqref{eq:fact} we need the
Fourier transform of the soft function defined according to
\begin{equation}
\label{eq:soft-ft}
W^{\{k\}}_{\{ab\}}(\omega,\mu)=
\int \frac{d z_0}{4 \pi}\,
e^{i \omega z_0/2} \,\hat W^{\{k\}}_{\{ab\}}(z_0,\vec 0,\mu).
\end{equation}
In~\eqref{eq:fact} we also decomposed the colour multi-indices in a set of  
basis structures $c^{(i)}_{\{a\}}$ defined below, and performed 
a projection on the irreducible representations $R_\alpha$ of the final state 
particle pair. Thus  $W^{R_\alpha}_{ii'}(\omega,\mu)$ is given by 
\begin{equation}
\label{eq:soft-coulomb}
W^{R_\alpha}_{ii'}(\omega,\mu)=  P^{R_\alpha}_{\{k\}}
c^{(i)}_{\{a\}} W^{\{k\}}_{\{ab\}}(\omega,\mu)c^{(i')*}_{\{b\}}.
\end{equation}
The remainder of the paper is concerned with the properties of the soft 
functions (\ref{eq:soft-general}) and (\ref{eq:soft-coulomb}). 
We remark that the soft function that appears in 
(\ref{eq:factor-thresh}) follows from contracting 
(\ref{eq:soft-general}) with the trivial colour factor 
$\delta_{k_1k_3}\delta_{k_2k_4}$ instead of the
projectors $ P^{R_\alpha}_{\{k\}}$ on the representations of the 
heavy-particle pair as
in~\eqref{eq:soft-coulomb}. Therefore the factorization of the Coulomb 
gluons results in a more complicated colour structure of the soft 
function than the original one, $W_{i i'}$, which is simply 
$\sum_{R_\alpha}W^{R_\alpha}_{ii'}$ due to the completeness of 
the projectors. For this reason, the original factorization formula 
(\ref{eq:factor-thresh}) should only be used for partonic thresholds 
where the relative velocities of the final state particles 
are relativistic, and Coulomb exchange is not enhanced. In this situation, 
the relevant soft function is (\ref{eq:soft-general}) with two 
unequal four-velocity 
vectors $v_1$, $v_2$ for the final state particles, 
contracted with $\delta_{k_1k_3}\delta_{k_2k_4}$. The main 
simplifications discussed in the following hold only for the 
soft function at threshold ($v_1=v_2$), and are thus 
applicable to the total 
partonic cross sections.

The paper is organized as follows. In section~\ref{sec:colour} we discuss the
colour structure of the soft function.  The simple kinematical structure of
the soft function $W^{R_\alpha}_{ii'}$ at threshold allows us to construct a
colour basis that diagonalizes the soft function to all orders of perturbation
theory. We shall see that
the four-particle soft function can be reduced to three-particle soft
functions corresponding to a single heavy particle in the final state.  In
section~\ref{sec:nnll}, after discussing the general structure of soft-gluon
and Coulomb resummation near threshold, we calculate the one-loop soft
function for general representations of initial and final state particles and
obtain the two-loop soft anomalous dimensions employing results
from~\cite{Becher:2009kw}.  This supplies all ingredients for NNLL
resummations for top-quark or sparticle pair production associated with the
(leading) soft function~\eqref{eq:soft-coulomb}. The two-loop soft anomalous
dimension at threshold exhibits Casimir scaling as has been found explicitly
at one-loop in previous
examples~\cite{Kidonakis:1997gm,Bonciani:1998vc,Kulesza:2008jb}.  Technical
results related to colour, Wilson lines and anomalous dimensions are
summarized in a number of appendices, including the colour bases for triplet,
anti-triplet and adjoint coloured particles in the initial and final state,
which covers the cases of interest.

%%%%%%%%%%%%%%%%%%%%%%%%%%%%%%%%%%%%%%%%%%%%%%%%%%%%%%%%%%%%%%%%%%%%%%%%%
\section{All-order colour structure of the soft function}
\label{sec:colour}

The physical picture of production of a heavy particle pair at
threshold suggests that soft gluon radiation cannot resolve the two
particles and couples to the total colour charge of the pair, determined by
the representation $R_\alpha$ in the decomposition of the product
representation $R\otimes R'$.  Therefore the structure of the (leading) 
soft function for the production of a non-relativistic particle pair in a 
given representation should be that of a single particle in the same 
representation. This is in agreement with the result that the one-loop soft
anomalous dimension at threshold is proportional to the quadratic Casimir
operator of the representation of the heavy-particle
pair~\cite{Kidonakis:1997gm,Bonciani:1998vc,Kulesza:2008jb}.  In this 
section we show how to obtain this structure, to all orders in 
perturbation theory, from the expression (\ref{eq:soft-coulomb}) 
for the soft function. In order to
accomplish this, we construct the projection operators $P^{R_\alpha}$ and the
basis tensors $c^{(i)}$ in the definition of the soft
function~\eqref{eq:soft-coulomb} from Clebsch-Gordan coefficients for the
decompositions of the product representations of the initial and final state
systems into irreducible ones.  
In this basis it is then easy to show that the
soft function is diagonal to all orders of perturbation theory.  Some
technical details of this construction are relegated to
appendix~\ref{app:colour-details}.  Explicit colour bases and projectors for
all production processes of heavy particles in the fundamental, antifundamental
and adjoint representations are provided in appendix~\ref{app:susy-colour}.
This covers all production processes of squarks and gluinos and completes
previous results for the colour bases of squark-antisquark and gluino
pair production~\cite{Kulesza:2008jb}.

\subsection{Notation}

With respect to colour, we use a notation that does
not distinguish particles and antiparticles. If, for example, $H'$ is
the anti-particle of $H$, then this convention implies that $H'$
transforms in the complex conjugate SU(3) representation of
$H$. Similarly, an initial-state antiquark transforms in the
anti-fundamental representation.  The Wilson lines $S$ inherit the 
corresponding representations. It is useful to perform a
decomposition of the product of the representations of the final
state and initial state particles into irreducible representations:
\begin{equation}
\label{eq:irreducible}
  r\otimes r' =\sum_{\alpha} r_\alpha\;,\qquad
R\otimes R'=\sum_{R_\alpha} R_{\alpha}.
\end{equation}
Examples relevant for the standard model and most extensions of the standard
model are particles in the fundamental and the adjoint representation where we
have the decompositions
 $3 \otimes \bar 3=1+8$ (e.g. top-antitop and
 squark-antisquark production),  
 $3 \otimes 3=\bar{3}+6$ (e.g. squark-squark
production), $3\otimes 8=3+\bar 6 +15$ (e.g. squark-gluino production) and $8
\otimes 8=1 \oplus 8_S \oplus 8_A \oplus 10 \oplus \overline{10} \oplus 27$
(gluino pair production). 
Generators of the SU(3) transformation in a representation $R$ are 
denoted by ${\bf T}^{(R)a}$. In practice we need the generators in the
fundamental and anti-fundamental representations, 
${\bf T}^{(3)a}_{a_1a_2}=T^a_{a_1a_2}$,  
${\bf T}^{(\bar 3)a}_{a_1a_2}=-T^a_{a_2a_1}$, and in the adjoint,   
${\bf T}^{(8)a}_{a_1a_2}=i f^{a_1aa_2}\equiv F^a_{a_1a_2}$. The
quadratic Casimir operator for a representation $R$ is denoted by 
$({\bf T}^{(R)a}{\bf T}^{(R)a})_{a_1a_2}=C_R\,\delta_{a_1a_2}$.

\subsection{Construction of projectors on irreducible representations}
\label{sec:colour-basis}

We now review a general method to construct the projection operators on the
irreducible representations of the heavy-particle pair appearing in the soft
function~\eqref{eq:soft-coulomb} from the  Clebsch-Gordan coefficients
that combine two objects transforming in
the representations $R$ and $R'$ of the group into a single object in an
irreducible representation $R_\alpha$.
The Clebsch-Gordan coefficients are defined as a unitary basis transformation,
\begin{equation}
\label{eq:clebsch-basis}
{\bf e}_{a_1 a_2} =\sum_{R_\alpha}
C^{R_\alpha}_{\alpha a_1a_2}  {\bf e}^{R_\alpha}_{\alpha},
\end{equation}
from the basis vectors of the tensor product space $R\otimes R'$ to basis
vectors of the irreducible representations $R_\alpha$ (see
e.g.~\cite{Tung:1985na}).
Here the ${\bf e}_{a_1 a_2}$ are elements of a complex, orthonormal basis of
the tensor product $R\otimes R'$ while the ${\bf e}^{R_\alpha}_{\alpha}$ are
elements of a basis of the irreducible representation $R_\alpha$. 
In this equation and in the following, repeated indices are summed over.
Unitarity of the basis transformation implies 
\begin{align}
\label{eq:unitary-clebsch}
\sum_{R_\alpha}C^{R_\alpha\ast}_{\alpha a_1a_2} C^{R_\alpha}_{\alpha b_1b_2}
&=\delta_{a_1b_1}\delta_{a_2b_2},\\
\label{eq:ortho_clebsch}
C^{R_\alpha\ast}_{\alpha a_1a_2}  C^{R_\beta}_{\beta a_1a_2}
&=\delta_{R_\alpha R_\beta}\delta_{\alpha\beta}.
\end{align}
A vector ${\bf V}$ in the tensor-product space can be written in the two bases
as ${\bf V}=V_{a_1a_2}{\bf
  e}_{a_1a_2}=\sum_{R_\alpha}V_\alpha {\bf e}_\alpha$, 
where the components are related by
\begin{align}
\label{eq:clebsch}
  V_{\alpha}&=C^{R_\alpha}_{\alpha a_1a_2}V_{a_1,a_2}, \\
\label{eq:inverse-clebsch}
  V_{a_1a_2}&=\sum_{R_\alpha}C^{R_\alpha\ast}_{\alpha a_1a_2} V_\alpha.
\end{align}
Consistency with the group transformations $V_\alpha\to
U_{\alpha\beta}^{(R)}V_\beta$ and the corresponding transformations of
$V_{a_1a_2}$ implies that the Clebsch-Gordan coefficients satisfy
\begin{equation}
\label{eq:clebsch-transform}
C^{R_\alpha}_{\alpha a_1a_2}U^{(R)}_{a_1b_1} U^{(R')}_{a_2b_2}=
U^{(R_\alpha)}_{\alpha\beta} C^{R_\alpha}_{\beta b_1b_2}.
\vspace*{0.2cm}
\end{equation}

We can now construct the  projectors $P^{R_{\alpha}}$ 
on the irreducible representations from the coefficients $C^{R_\alpha}$
according to
\begin{equation}
\label{eq:project-clebsch}
P^{R_\alpha}_{a_1a_2a_3a_4}=C^{R_\alpha\ast}_{\alpha a_1a_2}  
C^{R_\alpha}_{\alpha a_3a_4}.
\end{equation}
The orthonormalization condition~\eqref{eq:ortho_clebsch} implies that 
these are projectors satisfying
\begin{equation}
\label{eq:project}
  P^{R_{\alpha}}_{a_1a_2b_1b_2}P^{R_{\beta}}_{b_1b_2c_1c_2}=
\delta_{R_{\alpha} R_{\beta}}
P^{R_\alpha}_{a_1a_2c_1c_2},
\end{equation}
and the relation~\eqref{eq:unitary-clebsch} is equivalent to the completeness
relation
\begin{equation}
\label{eq:project-complete}
\sum_{R_{\alpha}}  P^{R_{\alpha}}_{a_1a_2b_1b_2}=
\delta_{a_1b_1}\delta_{a_2b_2}.
\end{equation}

As an example, consider final state particles in the $3$ and the $\bar
3$ representation of SU(3), e.g. top-antitop and squark-antisquark pairs.
The two particles combine to a singlet or octet, and the
Clebsch-Gordan coefficients read
\begin{equation}
\label{eq:cgk33}
  C^{(1)}_{a_1a_2}=\frac{1}{\sqrt{N_c}}\,\delta_{a_1a_2},\qquad
   C^{(8)}_{\alpha a_1a_2}=\sqrt 2 \,T^\alpha_{a_2a_1}.
\end{equation}
We obtain the familiar
projectors
\begin{equation}
\label{eq:projectors-fund}
 P^{(1)}_{\{a\}} =\frac{1}{N_c}\,\delta_{a_1a_2}\delta_{a_3a_4}, \qquad 
 P^{(8)}_{\{a\}} =2\,T^\alpha_{a_1a_2}T^\alpha_{a_4a_3}.
\end{equation}

\subsection{Colour basis for the hard production process}
\label{sec:colour-hard}

The colour structures $c_{\{a\}}^{(i)}$ in the
definition~\eqref{eq:soft-coulomb} describe the colour structure of
the hard-scattering process. They provide a decomposition of the
scattering amplitudes of the partonic process $pp'\to HH'$ into
independent basis tensors:
\begin{equation}
\label{eq:colour-amp}
\mathcal{A}_{pp'\{a\}}=\sum_i c^{(i)}_{\{a\}} \mathcal{A}_{pp'}^{(i)} .
\end{equation}
Here we use a multi-index notation 
$\{a\} = a_1a_2a_3a_4$. Repeated multi-indices
are summed over all four components.
We take the $c^{(i)}$ as an orthonormal basis satisfying
\begin{equation}
\label{eq:ortho-color}
c^{(i)}_{\{a\}} c^{(j) \ast}_{\{a\}} = \delta^{i j}.
\end{equation}      
The colour basis structures can be chosen as invariant tensors in the
representation $r\otimes r'\otimes \bar R\otimes\bar R'$:
\begin{equation}
\label{eq:global-basis}
c^{(i)}_{\{a\}}= U^{(R)\dagger}_{a_3b_3}  U^{(R')\dagger}_{a_4b_4} 
c^{(i)}_{\{b\}}  \, U^{(r)}_{b_1a_1} U^{(r')}_{b_2a_2}.
\end{equation}

A colour basis that is convenient for the discussion of the colour structure
of the soft function can be constructed from the Clebsch-Gordan coefficients,
similarly to the projectors.
Consider the subset of representations that appears in both the sets
$\{r_\alpha\}$ and $\{R_\alpha\}$ in~\eqref{eq:irreducible}, treating multiple
occurrences of equivalent representations in the decomposition as distinct.
We then form pairs $P_i=(r_{\alpha}, R_{\beta})$ of equivalent 
representations $r_\alpha$ and $R_\beta$, where the index 
$i$ enumerates the allowed
combinations. For example, in case of 
$8\otimes 8 \to 8\otimes 8$ we have the eight pairs
\begin{equation}
\label{eq:tensor-88}
P_i\in 
\{(1,1),\; (8_S,8_S),\;(8_A,8_S),\;(8_A,8_A),\;(8_S,8_A),\;\;(10,10),\;
(\overline{10},\overline{10}),\; (27,27) \}.
\end{equation}
The $10$ and $\overline{10}$ are inequivalent representations, 
so the pair $(10,\overline{10})$ is not allowed.
For the allowed pairs $P_i$, the colour structures 
\begin{equation}
\label{eq:prod-basis}
  c_{\{a\}}^{(i)}=\frac{1}{\sqrt{\text{dim}(r_\alpha)}}\,
  C^{r_\alpha}_{\alpha a_1a_2} C^{R_{\beta}\ast}_{\alpha a_3a_4}
\end{equation}
form an orthonormal basis satisfying~\eqref{eq:ortho-color}.  By
construction, the operators are invariant tensors under global SU(3)
transformations satisfying~\eqref{eq:global-basis}.  In
appendix~\ref{app:colour-details} we use colour conservation of the
scattering amplitude to show that the basis tensors always can be
chosen as in~\eqref{eq:prod-basis}.  

To illustrate this construction, we consider 
heavy particles in the $3$ and $\bar 3$ 
produced from a quark-antiquark initial state. There are only two 
possibilities to combine the initial and final state representations, 
$P_i=\{(1,1),(8,8)\}$. Using (\ref{eq:cgk33}) we obtain two operators 
for the basis of the hard-scattering process:
\begin{equation}
\label{eq:basis_33}
\begin{aligned}
c^{(1)}_{\{a\}} &= \frac{1}{N_c} \,\delta_{a_1a_2} \delta_{a_3a_4} ,&
c^{(2)}_{\{a\}} &=\frac{2}{\sqrt{D_A}} \,T^\beta_{a_2a_1} T^\beta_{a_3a_4}
\end{aligned}
\end{equation}
with $D_A=N_c^2-1$. This is the same basis found to diagonalize the
one-loop soft anomalous dimension at threshold~\cite{Kidonakis:1997gm}.
A complete list of all projectors and basis tensors for the production
of particles in the
fundamental and adjoint is given in appendix~\ref{app:susy-colour}.

\subsection{Diagonalization of the soft function}
\label{sec:colour-soft}
We now show that the soft function is diagonal to all orders of
perturbation theory in the basis constructed in
section~\ref{sec:colour-hard}. To achieve this, we express
the components of the soft function~\eqref{eq:soft-coulomb} in terms
of Wilson lines in the representations $R_\alpha$, as suggested by
the physical picture of soft radiation off the total colour charge of
the final state system. We first note that the Wilson lines satisfy a
relation analogous to~\eqref{eq:clebsch-transform},
\begin{equation}
\label{eq:combine-wilson}
C^{R_\alpha}_{\alpha a_1a_2}S^{(R)}_{v,a_1b_1}
S^{(R')}_{v,a_2b_2}=S^{(R_\alpha)}_{v,\alpha\beta}
C^{R_\alpha}_{\beta b_1b_2}.
\end{equation}
A proof of this relation is given in
appendix~\ref{app:colour-details}.  Using this identity and the
completeness relation~\eqref{eq:unitary-clebsch} of the Clebsch-Gordan
coefficients we can write the soft function~\eqref{eq:soft-general} in
terms of Wilson lines for single particles in the irreducible
representations $R_\alpha$:
\begin{eqnarray}
&&\hat W^{\{k\}}_{\{ab\}}(z,\mu)
=\sum_{R_\alpha,R_\beta}C^{R_\beta\ast}_{\beta b_3b_4}  
C^{R_\beta}_{\kappa k_1k_2} 
C^{R_\alpha}_{\alpha a_3a_4} C^{R_\alpha\ast}_{\lambda ,k_3k_4} 
\nonumber \\
&&\hspace*{4cm}
\times \,\langle 0|\overline{\mbox{T}}[S^{R_\beta}_{v,\beta \kappa} 
S^\dagger_{\bar{n},jb_2}S^\dagger_{n, ib_1} ](z)
\mbox{T}[ S_{n,a_1i} S_{\bar{n},a_2j} 
S^{R_\alpha\dagger}_{v,\lambda \alpha} ](0)
|0\rangle.
\qquad
\label{eq:combine-soft}
\end{eqnarray}
This combination of the two Wilson lines in the representations $R$
and $R'$ to a single one in the representations $R_\alpha$
is only possible if the heavy particle pair is produced close to threshold
and not for generic kinematics where Wilson lines in different 
directions $v_1,v_2$ appear in the soft function.

In~\eqref{eq:combine-soft} the Wilson lines related to the final state
system are still in two different representations. This structure
simplifies when we compute the components of the soft
function~\eqref{eq:soft-coulomb} where the soft function is contracted
with a projector on an irreducible representation. Using the
definition of the projectors~\eqref{eq:project-clebsch} we have the
identity
\begin{equation}
C^{R_\beta}_{\kappa,k_1k_2} P^{R_\alpha}_{\{k\}}
C^{R_\gamma\ast}_{\lambda,k_3k_4} 
=\delta_{\kappa\lambda} \delta_{R_\alpha R_\gamma}\delta_{R_\alpha R_\beta},
\end{equation}
i.e. the projectors enforce that the representations $R_\beta$ and $R_\gamma$
are identical. For distinct equivalent representations such as the $8_S$ and
$8_A$ representations in the decomposition of $8\otimes 8$ we obtain a
vanishing result.  We can then express the components of the soft
function~\eqref{eq:soft-coulomb} as
\begin{equation}
\label{eq:soft-coulomb-2}
 W^{R_\alpha}_{ii'}(\omega,\mu)=  
c^{R_\alpha(i)}_{\{a\alpha\}} W^{R_\alpha}_{\{a\alpha,b\beta\}}(\omega,\mu)
c^{R_\alpha(i')\ast }_{\{b\beta\}},
\end{equation}
where we have introduced  the  soft function for  the production of a
\emph{single particle} in the representation $R_\alpha$
\begin{equation}
\label{eq:soft-R}
\hat W^{R_\alpha}_{\{a\alpha ,b\beta \}}(z,\mu)\equiv
\langle 0|\overline{\mbox{T}}[S^{R_\alpha}_{v,\beta \kappa}
S^\dagger_{\bar{n},jb_2} S^\dagger_{n, ib_1} ](z)
\mbox{T}[S_{n,a_1i}S_{\bar{n},a_2j}S^{R_\alpha\dagger}_{v,\kappa\alpha}](0)|
0\rangle .
\end{equation}
Here we have extended our multi-index convention to the indices of the
irreducible representations of the final state system by defining
$\{a\alpha\}=a_1a_2\alpha$.  Note that the index $\kappa$ of the two
final state Wilson lines in~\eqref{eq:soft-R} is contracted, so this
is analogous to the soft function appearing in
the conventional treatment where Coulomb gluons are not
factorized. This function has been considered in~\cite{Idilbi:2009cc}
for the case of a single colour-octet scalar.  The colour basis
tensors of the production operators in the new notation are given by
\begin{equation}
\label{eq:c-ralpha}
c^{R_\alpha(i)}_{\{a\alpha\}}\equiv c^{(i)}_{\{a\}}C^{R_\alpha}_{\alpha a_3a_4}
=\frac{1}{\sqrt{\text{dim}(r_\alpha)}}\,
  C^{r_\alpha}_{\alpha a_1a_2}\delta_{R_\alpha R_{\beta}}.
\end{equation}
As indicated, they are nonvanishing only if the final state
representation $R_\alpha$ is identical to the final state
representation $R_\beta$ in the pair
$P_i=(r_\alpha,R_{\beta})$ that defines the tensor $c^{(i)}$. This 
implies that $r_\alpha$ and $R_\alpha$ must 
be equivalent, as is intuitively clear.

\begin{figure}[t]
 \begin{center}
 \hskip-1cm
  \includegraphics[width=0.6\textwidth]{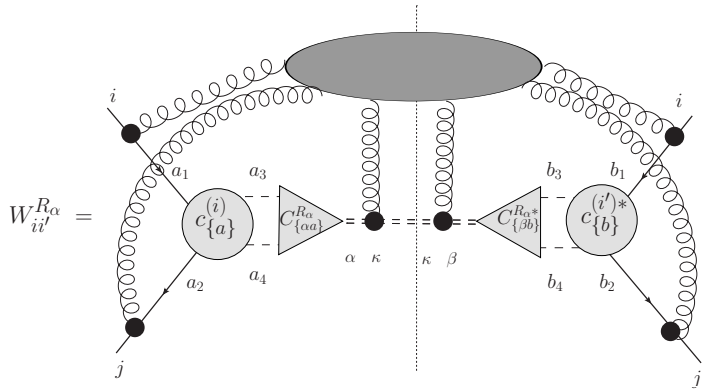}  
 \caption{Structure of the simplified expressions of the soft
   function~\eqref{eq:soft-coulomb-2} and~\eqref{eq:soft-R}.}
 \label{fig:soft-coulomb}
\end{center}
\end{figure}

In~\eqref{eq:soft-R} we have reduced the problem of soft
gluon radiation in pair production of heavy particles to that of the
production of a single particle in the representation $R_\alpha$. This
is the first main result of this paper and will simplify the
computation of the one-loop soft function and the 
two-loop soft anomalous dimension.  The structure of the
results~\eqref{eq:soft-coulomb-2} and~\eqref{eq:soft-R} is sketched in
figure~\ref{fig:soft-coulomb}.

Eq.~\eqref{eq:c-ralpha} for the basis tensors automatically
leads to the vanishing of a large number of components of the soft
function.  The component $W^{R_\alpha}_{ii'}$ is nonzero only if the
representation $R_\alpha$ of the final state system is
\emph{identical} to both final state representations $R_{\beta}$ and
$R_{\beta '}$ and hence, by construction of the basis tensors,
\emph{equivalent} to both initial state representations $r_{\alpha}$
and $r_{\alpha'}$ in the pairs $P_i=(r_\alpha,R_\beta)$ and
$P_{i'}=(r_{\alpha'},R_{\beta'})$.  Therefore the soft function is
block-diagonal in the basis~\eqref{eq:prod-basis} with off-diagonal
elements arising only if several representations in the
decomposition~\eqref{eq:irreducible} of the initial state system are
equivalent.  (A non-trivial structure does not appear if the several
equivalent representations appear only in the final state since the
projectors $P^{R_\alpha}$ always project on a unique representation.)
Since in the decompositions of $3\otimes 3$, $3\otimes \bar 3$ and
$3\otimes \bar 8$ no representation occurs more than once, the
matrices $W^{R_\alpha}_{ii'}$ are diagonal if at least one quark or
anti-quark is present in the initial state.  The only example where a
non-trivial matrix structure can arise is the gluon-gluon channel.

For the example of an $8\otimes 8$ final state produced from
gluon-gluon fusion, according to~\eqref{eq:tensor-88}, only the soft
functions for the two octet final states $8_S$ and $8_A$ are
(potentially) non-trivial two-by-two matrices,
\begin{equation}
\label{eq:soft-block}
{\bf W}^{8_S}=\begin{pmatrix}
0& \cdots&\cdots &\cdots\\
\vdots &W^{8_S}_{22}&W^{8_S}_{23}&\\
\vdots & W^{8_S}_{32}&W^{8_S}_{33}&\\
\vdots&&&0&\\
&&&&\ddots
\end{pmatrix}\;,\qquad
{\bf W}^{8_A}=\begin{pmatrix}
0& \cdots&\cdots &\cdots\\
\vdots&\ddots&&&\\
\vdots& &W^{8_A}_{44}&W^{8_A}_{45}&\\
\vdots& &W^{8_A}_{54}&W^{8_A}_{55}\\
&&&&\ddots
\end{pmatrix}
\end{equation}
with entries in the order of the basis elements 
used in~\eqref{eq:tensor-88}. The soft functions for the 
$1$, $10$, $\overline{10}$ and $27$ representations, however, 
consist of a single non-vanishing element. Analogously,
for the production of a $3\otimes 3$ final state from gluon-gluon fusion the
singlet soft function consists of a single non-vanishing entry while 
the colour-octet soft
function contains a potentially non-trivial two-by-two submatrix mixing $8_A$
and $8_S$ initial states.

 We now use Bose symmetry of the soft function to show that the
 off-diagonal elements in the soft function for the production of a
 colour-octet state vanish, so that the matrices~\eqref{eq:soft-block}
 and the analogous ones for a $3\otimes 3$ final state are in fact
 diagonal to all orders in perturbation theory.  First note that,
 independent of the nature of the final state system, these
 off-diagonal matrix elements involve a combination of Wilson lines
 contracted with the Clebsch-Gordan coefficients $C^{(8_A)}_{\alpha
   a_1a_2}$ and $C^{(8_S)}_{\alpha a_1a_2}$ defined
 in~\eqref{eq:cgk83}:
\begin{equation}
\label{eq:off-diag-soft}
C^{(8_S)\ast}_{\beta b_1b_2}[S^\dagger_{n, ib_1} 
S^\dagger_{\bar{n},jb_2} ](z)[ S_{\bar{n},a_2j} S_{n,a_1i}](0)
C^{(8_A)}_{\alpha a_1a_2}
 \propto [S^\dagger_{n, ib_1} D^{\beta}_{ b_1b_2}
 S^\dagger_{\bar{n},jb_2} ](z)[ S_{\bar{n},a_2j} F^{\alpha}_{ a_2a_1}
 S_{n,a_1i}](0).
\end{equation}
Because the two incoming Wilson lines are indistinguishable, 
the soft function must be invariant under the exchange of their colour 
labels and momenta, i.e. the exchange
$(n,i) \leftrightarrow (\bar{n},j)$.
This statement translates into the following equation:
\begin{eqnarray}
&& \lbrack S^\dagger_{n, ib_1} D^{\beta}_{ b_1b_2}
S^\dagger_{\bar{n},jb_2} \rbrack(z)
[S_{\bar{n},a_2j} F^{\alpha}_{ a_2a_1} S_{n,a_1i}](0)
\nonumber\\
&&\hspace*{1cm} = \,
[S^\dagger_{\bar n, jb_1}D^{\beta}_{ b_1b_2}S^\dagger_{n,ib_2}](z)
[S_{n,a_2i} F^{\alpha}_{ a_2a_1} S_{\bar n,a_1j}](0)
\nonumber\\
&&\hspace*{1cm} = \,
[S^\dagger_{n, ib_1} D^{\beta}_{b_2b_1} S^\dagger_{\bar n,jb_2}](z) 
[S_{\bar n,a_2j} F^{\alpha}_{ a_1a_2} S_{\bar n,a_1i}](0)
\nonumber\\
&&\hspace*{1cm} =\, 
-[S^\dagger_{n, ib_1} D^{\beta}_{ b_1b_2}
S^\dagger_{\bar{n},jb_2}](z)[S_{\bar{n},a_2j} 
F^{\alpha}_{ a_2a_1} S_{n,a_1i}](0)
\nonumber\\
&&\hspace*{1cm} =\, 0.
\end{eqnarray}
Here we have used the symmetry properties of the $D$ and $F$ tensors.
Therefore the off-diagonal terms of the colour-octet soft function
such as in~\eqref{eq:soft-block} vanish.
Furthermore from the explicit results in section~\ref{sec:nnll} we 
deduce that the one-loop soft function and two-loop soft anomalous 
dimensions are determined by the quadratic Casimir operators, so 
the diagonal elements of $W^{8_A}$ and $W^{8_S}$ are all identical.  
This would extend to higher orders if Casimir scaling held to all
orders.  However, presently we do not have a proof for this.

To summarize we have shown that for all initial states relevant to hadron
colliders the soft function is diagonal to all orders in a colour basis for
the hard-scattering process given by~\eqref{eq:prod-basis}.  This holds
independent of the nature of the final state particles, i.e.~equally for
top-quarks, squarks or gluinos.  As mentioned at the end of
section~\ref{sec:introduction}, the approach used in most phenomenological
applications corresponds to using the soft function $W_{ii'}$ obtained by
summing up all final state representations so our results apply to this case
as well.  Let us briefly recall the main ingredients used in order to
arrive at this result:
\begin{itemize}
\item The Coulomb interaction is diagonalized by the
  decomposition of the final state system into irreducible
  representations~\eqref{eq:irreducible}, leading to the definition of the
  components of the soft function~\eqref{eq:soft-coulomb} where the
  Wilson lines associated to the
  final-state particles are projected onto the irreducible representations.
\item The hard-scattering amplitudes are
  colour conserving, eq.~\eqref{eq:global-basis}, which allowed us to
  choose the basis of colour tensors according
  to~\eqref{eq:prod-basis}.
\item For a heavy particle pair produced directly at threshold both
  particles have the same velocity, allowing to combine the two 
  final-state Wilson lines into a single one.
\item Due to Bose symmetry of the soft function there is no
  interference of the production from a symmetric and antisymmetric
  colour octet.
\end{itemize}

\subsection{Examples}
\label{sec:examples}
In this subsection we give some examples of the formalism in order 
to show how it  is related to the
colour bases used in previous computations. 
As an illustration of  the result~\eqref{eq:soft-coulomb-2} for the
components of the soft function, we give explicit expressions for the
example of a $3\otimes \bar 3$ final state and a quark-antiquark or
gluon-gluon initial state. In subsection~\ref{sec:example-octet} we compare
the basis for an  $8\otimes 8$ final state produced in gluon fusion to that
used previously~\cite{Kulesza:2008jb}.

\subsubsection{Soft function for a $3\otimes \bar 3$ final state}

For quark-antiquark initiated production of a heavy particle pair in the 
$3\otimes\bar 3$ representation, the required colour basis is
given by~\eqref{eq:basis_33} so the soft functions for the singlet and octet
final states, $W^1_{ii'}$ and $W^8_{ii'}$, are two-by-two matrices.
 The only non-vanishing component of the
soft function for the singlet case is identical to that in Drell-Yan
production~\cite{Korchemsky:1993uz,Becher:2007ty}
\begin{equation}
\label{eq:soft-dy}
  \hat W^{1}_{11}(z,\mu) =
\frac{1}{N_c}\langle 0| \mbox{Tr}[
\overline{\mbox{T}}[S^\dagger_{n} S_{\bar{n}}](z)
\mbox{T}[S_{\bar{n}}^\dagger S_{n}](0)]|0\rangle
=\hat W_{DY}(z,\mu).
\end{equation}
Here we have deviated from our usual notation of using the
anti-fundamental representation for anti-particles and expressed the
result using only Wilson lines in the fundamental representation in
order to simplify the matrix structure.
 Similarly, the only
non-vanishing component of the colour-octet soft function is given by
\begin{equation}
\hat W^{8}_{22}(z,\mu)=\frac{2}{N_c^2-1}\,
 \langle 0| \mbox{Tr}[
\overline{\mbox{T}}[S_{v,bc}S^\dagger_n T^{b} S_{\bar{n}}](z)
\mbox{T}[S^{\dagger}_{\bar{n}} T^a S_n S^\dagger_{v,ca}](0)] |0\rangle ,
\end{equation}
where the $S_v$ are in the adjoint representation and the trace is over the
fundamental representation.
 
For the production of a $3\otimes \bar 3$ final state from gluon
fusion there are three possible combinations of initial and final 
state representations: 
\begin{equation}
P_i\in\{(1,1),\; (8_S,8),\;(8_A,8)\}.
\end{equation}
The Clebsch-Gordan coefficients and colour-basis elements for this case are
collected in appendix~\eqref{eq:3-3b}.  Since the set of $P_i$'s has
three elements, the singlet and octet soft
functions are three-by-three matrices. The only non-vanishing element of
the soft function for the singlet channel is, up to normalization, again given
by~\eqref{eq:soft-dy}, where now the Wilson lines $S_{\bar{n}}$ and $S_{n}$
are in the adjoint representation.  An octet final state can be produced
either from a symmetric or antisymmetric octet initial state corresponding to
the basis elements $c^{(2)}$ and $c^{(3)}$ in~\eqref{eq:basis_83}.  The
non-vanishing diagonal elements of the soft-function matrix for the octet
channel in this basis are given by
\begin{align}
\hat W^{8}_{22}(z,\mu)&=\frac{N_c}{(N_c^2-1)(N_c^2-4)}\,
 \langle 0| \mbox{Tr}[
\overline{\mbox{T}}[S_{v,ac}S^\dagger_n D^{a} S_{\bar{n}}](z)
\mbox{T}[S^{\dagger}_{\bar{n}} D^b S_n S^\dagger_{v,cb}](0)] |0\rangle, \\
\hat W^{8}_{33}(z,\mu)&=\frac{1}{N_c(N_c^2-1)}\,
 \langle 0| \mbox{Tr}[
\overline{\mbox{T}}[S_{v,bc}S^\dagger_n F^{b} S_{\bar{n}}](z)
\mbox{T}[S^{\dagger}_{\bar{n}} F^a S_n S^\dagger_{v,ca}](0)] |0\rangle .
\end{align}
Here we have used the fact that the Wilson lines in the adjoint,
$S_n,S_{\bar n}$, are real to write the matrix product in a convenient
form.  The expressions for the octet soft functions agree precisely
with~\cite{Idilbi:2009cc} (up to their notation for the result in
momentum space).  The off-diagonal elements $W^8_{23}$ and $W^8_{32}$
involve the structure~\eqref{eq:off-diag-soft} and therefore vanish by
symmetry arguments.

\subsubsection{Colour octet states in $8\otimes 8\to 8\otimes 8$ }
\label{sec:example-octet}

We would like to comment briefly on previous results for the basis for
the gluon-induced production of heavy particles in the adjoint
representation (e.g. gluinos).  The complete basis for this case is
given in appendix~\ref{app:susy-colour}.  Here we will only need the
four operators corresponding to the different combinations of $8_S$
and $8_A$ in the initial and final state~\eqref{eq:basis-88-octet}.
This basis differs slightly from the one constructed
in~\cite{Kidonakis:1998nf} for dijet production where the linear
combinations $\frac{1}{\sqrt 2}(c^{(3)}\pm c^{(5)})$ have been used.
In~\cite{Kulesza:2008jb} that basis has been shown to diagonalize the
one-loop soft anomalous-dimension matrix corresponding to the soft
function $W_ {ii'}$ discussed below~\eqref{eq:soft-coulomb} at
threshold, i.e. for $v_1=v_2$.  From the general arguments given above
and from an explicit calculation we find that the one-loop soft
functions for colour octet final states $W^{8_S}_{ii'}$ and
$W^{8_A}_{ii'}$ are diagonal in the basis~\eqref{eq:basis-88-octet}
but not in the one used in~\cite{Kulesza:2008jb}. However, since the
Coulomb functions $J_{8_A}$ and $J_{8_S}$ are identical, only the sum
$W^{8_S}+W^{8_A}$ enters the cross section~\eqref{eq:fact}. Since the
 off-diagonal terms cancel in the sum, our result is consistent with
the one in~\cite{Kulesza:2008jb}.  Similar remarks apply to the colour tensors
related to the $10$ and $\overline{10}$ where the basis used
in~\cite{Kidonakis:1998nf,Kulesza:2008jb} is appropriate for the sum
$W^{10}+W^{\overline{10}}$ that is relevant to the cross section.

\section{Ingredients for NNLL threshold resummation}
\label{sec:nnll}

The resummation of threshold logarithms proceeds by using the
factorization scale independence of the total cross section to derive
renormalization group equations for the hard function $H_{ii'}$, which
appears in eq.~(\ref{eq:fact}), and the soft function. To define 
the NLL, NNLL, etc. approximations, we note that near threshold 
the usual expansion, where $\alpha_s \ln\beta$ counts as order one, 
is combined with an expansion in $\beta$, such that $\alpha_s/\beta$ 
also counts as one. This leads to a parametric representation 
of the expansion of the cross section in the form
\begin{eqnarray}
\label{eq:syst}
\hat{\sigma}_{p p'} &=& \,\hat \sigma^{(0)}\, 
\sum_{k=0} \left(\frac{\alpha_s}{\beta}\right)^k \,
\exp\Big[\underbrace{\ln\beta\,g_0(\alpha_s\ln\beta)}_{\mbox{(LL)}}+ 
\underbrace{g_1(\alpha_s\ln\beta)}_{\mbox{(NLL)}}+
\underbrace{\alpha_s g_2(\alpha_s\ln\beta)}_{\mbox{(NNLL)}}+\ldots\Big]
\nonumber\\[0.2cm]
&& \,\times
\left\{1\,\mbox{(LL,NLL)}; \alpha_s,\beta \,\mbox{(NNLL)}; 
\alpha_s^2,\alpha_s\beta,\beta^2 \,\mbox{(NNNLL)};
\ldots\right\}, 
\end{eqnarray}
which reproduces the standard structure~\cite{Bonciani:1998vc} away from
threshold for $k=0$ and no expansion in $\beta$. Thus, in fixed orders, LL
includes relative to the tree term $\hat \sigma^{(0)}$ all terms of the form
\begin{equation}
\mbox{LL} \qquad 
\alpha_s \left\{\frac{1}{\beta},\ln^2\beta\right\}; 
\,\alpha_s^2 \left\{\frac{1}{\beta^2},\frac{\ln^2\beta}{\beta},
\ln^4\beta \right\};
\ldots,
\end{equation}
while NLL and NNLL further include all terms
\begin{eqnarray}
\mbox{NLL}&& \qquad 
\alpha_s \ln\beta; 
\,\alpha_s^2 \left\{\frac{\ln\beta}{\beta},
\ln^3 \beta \right\};
\ldots,
\nonumber\\
\mbox{NNLL}&& \qquad 
\alpha_s \left\{1, \beta \times \ln^{2,1} \beta 
\right\}; 
\,\alpha_s^2 \left\{\frac{1}{\beta},
\ln^{2,1} \beta , \beta \times \ln^{4,3} \beta \right\};
\ldots,
\end{eqnarray}
respectively. Note that while the LL approximation sums soft
logarithms of the form $\alpha_s^n\log
\beta^m=\alpha_s\log\beta^2,\dots$ with $n+1\leq m\leq 2n$, this does
not include all terms of this form at $\mathcal{O}(\alpha_s^2)$.
Similarly, the NLL approximation sums soft logarithms of the form $(\alpha_s\log \beta)^n$ but $\mathcal{O}(\alpha_s^2\log\beta^2)$ terms from the
interference of the one-loop hard function and leading soft logarithms
are included only at NNLL. The NNLL terms proportional to $\beta$ 
originate from $\beta$-suppressed corrections to the hard functions 
and the soft gluon couplings. The former, however, vanish, since 
S-wave and P-wave production processes are not interfering. The 
$\beta$-suppressed soft corrections average to zero in the total cross section 
at $\mathcal{O}(\alpha_s)$ (and probably as well in 
higher orders), such that no terms of the form 
$\alpha_s \beta \times \ln^{2,1} \beta$ are present in the fixed-order 
expansion.

We would now like to explain briefly the several functions appearing in the
factorization formula~\eqref{eq:fact}. We also discuss how the
expansion~\eqref{eq:syst} is generated from this expression and, starting from
NNLL, additional contributions of a similar factorized form.  The derivation
of the factorization formula (\ref{eq:fact}) in \cite{BFS} relies on
soft-collinear and potential non-relativistic effective field theory, and the
fact that soft gluon interactions can be decoupled from collinear and
potential fields in the leading-order effective Lagrangian using
field redefinitions involving the Wilson lines~\eqref{eq:def-wilson-coll}
and~\eqref{eq:def-wilson}.  For the initial-state partons, this redefinition
is identical to that in the derivation of the factorization formula for the
Drell-Yan process at partonic threshold~\cite{Becher:2007ty}. 
The redefinitions  for the final state particles are a generalization of 
those used for non-relativistic $W$-bosons~\cite{Falgari:2009zz} to 
the case of a colour Coulomb force. The soft
function~\eqref{eq:soft-general} collects the Wilson lines arising from these
field redefinitions. The function $J_{R_\alpha}$ factorizes potential effects
and sums Coulomb gluon exchange related to the attractive or repulsive Coulomb
force in the irreducible colour representations $R_\alpha$. It is defined as a
correlation function of non-relativistic fields and can be expressed in terms
of the imaginary part of the zero-distance Coulomb Green function of the
Schr{\"o}dinger equation. The hard function is defined in terms of
squared short-distance coefficients, $H_{ii'}\propto
C_{pp'}^{(i)}C_{pp'}^{(i')\ast}$, analogous to the corresponding
treatments of heavy-particle pair production in $e^+ e^-$ 
collisions~\cite{Beneke:1999qg} 
and of the Drell-Yan process~\cite{Becher:2007ty}. The
coefficients $C_{pp'}^{(i)}(M,\mu)$ encode the contribution of hard
momenta to the process $pp'\to HH'$ and are obtained from the
scattering amplitude for the partonic subprocess evaluated directly at
threshold~\cite{BFS}.

For resummation at NLL accuracy, the required ingredients in 
eq.~(\ref{eq:fact}) are the hard function $H_{i i'}$ and the soft
function $W_{i i'}^{R_\alpha}$ both at tree-level and 
the one-loop anomalous dimensions appearing
in the evolution equations, with the exception of the so-called cusp
anomalous dimension related to the leading logarithms that is required
at two loops. We note that part of 
the NLL $\alpha_s^2/\beta\times \ln\beta$ term
arises from the running coupling in the Coulomb potential and is correctly
taken into account by choosing the scale to be $M\beta$ in $J_{R_\alpha}$.

For resummation at NNLL accuracy the hard and soft
function are needed at one-loop level, the cusp anomalous dimension at
three loops and all remaining anomalous dimensions at two loops.
While most of the anomalous dimensions are known to the required order
or higher from studies of deep-inelastic scattering, the Drell-Yan
process or Higgs production from gluon fusion, the soft anomalous
dimension for pair production of heavy coloured particles is currently
available only at
one-loop~\cite{Kidonakis:1997gm,Bonciani:1998vc,Kulesza:2008jb},
despite recent progress on massive amplitudes at the two-loop
level~\cite{Kidonakis:2009ev,Mitov:2009sv,Becher:2009kw}.

In addition to the soft corrections considered in this paper, further
logarithmic contributions arise from higher-order terms in the effective
Lagrangian or the production operators where soft gluons do not
decouple after the field redefinitions.  An example is the dipole
interaction $\vec x\cdot \vec E$ in the potential non-relativistic QCD
Lagrangian.  These terms, however, can be treated as perturbations so
that the entire expansion (\ref{eq:syst}) can be constructed as a sum
of terms in the factorized form $H^{(k)}\times J^{(k)}\star W^{(k)}$
with new hard, potential and soft functions, of which
eq.~(\ref{eq:fact}) constitutes only the leading term in the expansion
in $\beta$. Due to the $\beta$ suppression, the first correction to
the leading term arises at NNLL.  Therefore further NNLL terms in
addition to~\eqref{eq:fact} may arise from two sources related to
higher-dimensional terms in the expansion in $\beta$ in the
factorization formula: first, corrections to the Coulomb function
$J_{R_\alpha}$ due to subleading heavy-quark potentials contribute
NNLL terms, which at ${\cal O}(\alpha_s^2)$ are of the form
$\alpha_s^2\,\log\beta$~\cite{Beneke:1999qg}.  Second,
higher-dimensional soft functions with insertions of the $\vec{x}\cdot
\vec{E}$ interaction potentially also contribute
$\alpha_s^2\,\log^{2,1}\beta$ terms from the interference of a
$\beta$-suppressed soft-gluon emission with one-Coulomb exchange.  As
mentioned before, there are no corrections linear in $\beta$ related
to the hard function.
 Finally, we mention that before the
convolution with the parton distributions, the NNLL resummed cross
section should be matched to a fixed-order two-loop calculation that
is not yet available for the processes of interest.

In this section we continue our investigation of the (leading) soft
function and provide all those ingredients required for an NNLL
resummation for pair production of arbitrary coloured particles that are
related to (\ref{eq:fact}) except for the process-dependent one-loop
hard functions $H_{ii'}$, i.e.~we provide the one-loop soft function
$W_{i i'}^{R_\alpha}$ and its two-loop soft anomalous dimension.  In
subsection~\ref{sec:soft-one} we compute the one-loop soft function
for initial-state particles in arbitrary representations $r$ and $r'$
of SU(3) and a final state system in an arbitrary representation
$R_\alpha$, using the reduction of the general soft function for two
final state Wilson lines~\eqref{eq:soft-general} to the soft
function~\eqref{eq:soft-R} with a single Wilson line representing the
final state system, achieved in section~\ref{sec:colour}.  In
subsection~\ref{sec:gamma} we then obtain the two-loop soft anomalous
dimension using results
of~\cite{Becher:2009kw,Korchemsky:1991zp,Kidonakis:2009ev}.
In subsection~\ref{sec:compare} we relate our results to 
the conventions used in the NNLL treatment of the top-quark production 
cross section in Mellin space~\cite{Moch:2008qy}.

\subsection{One-loop soft function}
\label{sec:soft-one}
In this subsection we compute the one-loop term in the loop expansion
of the soft function
\begin{equation}
\label{eq:soft-alpha}
  \hat W_{\{a\alpha ,b\beta \}}^{R_\alpha}(z_0,\mu)=\sum_{n=0}^\infty
 \left(\frac{\alpha_s(\mu)}{4\pi}\right)^{n}  
\hat W_{\{a\alpha ,b\beta \}}^{(n)R_\alpha}(z_0,\mu).
\end{equation}
The $n$-loop contribution to the soft function is obtained from the
definition~\eqref{eq:soft-R} by expanding each Wilson line up to
order $g_s^{2n}$ and keeping all contributions to the soft function of order
$g_s^{2n}$.  At tree level, the soft function in position and momentum
space is simply given by
\begin{equation}
\begin{aligned}
 \hat W_{\{a\alpha ,b\beta \}}^{R_\alpha(0)}(z_0,\mu)&=
\delta_{a_1 b_1}\delta_{a_2 b_2}\delta_{\alpha\beta},\\
  W_{\{a\alpha ,b\beta \}}^{R_\alpha(0)}(\omega,\mu)&=
\delta_{a_1 b_1}\delta_{a_2 b_2}\delta_{\alpha\beta}\; \delta(\omega).
\end{aligned}
\end{equation}

\begin{figure}[t]
  \begin{center}
    \includegraphics[width=\textwidth]{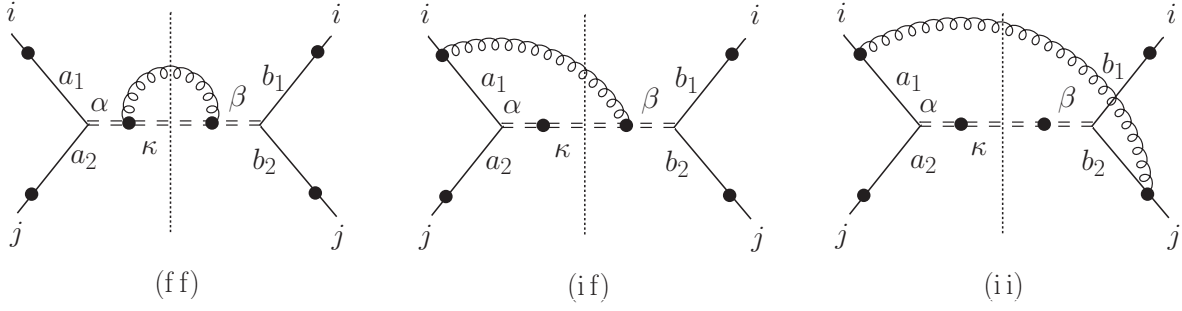}
  \caption{Examples for diagrams contributing to the soft function}
  \label{fig:soft-one}
  \end{center}
\end{figure}

At the one-loop order, the expansion of the Wilson lines gives rise to
real and virtual initial-initial~(ii), initial-final~(if) and
final-final~(ff) state interference diagrams. Examples for diagrams
contributing to the real corrections are shown in
figure~\ref{fig:soft-one}.  As an example consider the initial-final
state interference diagram denoted by $(\text{if})$ in
figure~\ref{fig:soft-one}. This arises from the contribution to the
soft function~\eqref{eq:soft-R} where the Wilson lines $S^{(r)}_{n,a_1
  i}(0)$ and $S^{(R_{\alpha})}_{v,\beta\kappa}(z)$ both contribute at 
order $g_s$ while all the other Wilson lines give trivial contributions. 
This diagram arises from the expectation value
\begin{align}
\label{eq:if-example}
&\langle 0|\overline{\mbox{T}}[S^{(R_\alpha)}_{v,\beta \kappa}(z)]
\mbox{T}[ S^{(r)}_{n,a_1i}(0)]|0\rangle |_{\text{if}}
\nonumber\\
&=\left(-ig_s {\bf T}^{(R_\alpha)c}_{\beta\kappa}\right)
\left(ig_s{\bf T}^{(r)d}_{a_1 i}\right)
\int_0^{\infty}\!\!\!\!ds \int_{-\infty}^0\!\!\!\!dt\; 
\langle 0|\overline{\mbox{T}}[v\cdot A^c(v(z^0+s))]
\mbox{T}[n\cdot A^d(tn)]|0\rangle |_{\text{if}} \nonumber\\
&=-4\pi\alpha_s
 {\bf T}^{(R_\alpha)c}_{\beta\kappa}{\bf T}^{(r)c}_{a_1i}
\;
(v\cdot n) \int_{-\infty}^0\!\!\!\!ds \int_{-\infty}^0\!\!\!\!dt\; 
D^+(v(z^0-s)-nt).
\end{align}
Here we have introduced the cut gluon propagator in position space
\begin{equation}
  D^{ab,+}_{\mu\nu}(x)
=\langle 0|A^a_\mu(x) A^b_{\nu}(0)|0\rangle
\equiv(-g_{\mu\nu})\delta^{ab}D^+(x).
\end{equation}
It is convenient to evaluate the integrals arising in the soft
function directly in position space using the dimensionally
regularized form of the cut-gluon propagator~\cite{Korchemsky:1992xv}
\begin{equation}
D^+(x)=
 \frac{\Gamma(1-\epsilon)}{4\pi^{2-\epsilon}}
\frac{1}{[-(x_+-i \delta)(x_--i\delta)]^{1-\epsilon}}
\end{equation}
with $x_+=n\cdot x$ and $x_-=\bar n\cdot x$, and   
where the last expression holds for 
$x_\perp^\mu\equiv x^\mu-x_- n^\mu/2 -x_+ \bar n^\mu/2=0$.
The usual Feynman propagator in position space is given by
\begin{equation}
  D^{ab}_{\mu\nu}(x)=(-g_{\mu\nu})\delta^{ab}D(x)
=(-g_{\mu\nu})\delta^{ab}
 \frac{\Gamma(1-\epsilon)}{4\pi^{2-\epsilon}}
\frac{1}{(-x^2+i \delta)^{1-\epsilon}}.
\end{equation}

Proceeding in the same way for all diagrams we obtain the one-loop 
soft function. Setting scaleless integrals to zero, only the real 
corrections are non-vanishing. The complete one-loop soft function 
can then be written in terms of 
group-theory factors $C$ and integrals $\mathcal{I}$ as
\begin{equation}
\label{eq:soft-result-1}
  \hat W^{(1)\,R_\alpha}_{\{a\alpha ,b\beta \}}(z_0,\mu) = -(4\pi)^2\left[
C^{(\text{ii})}_{\{a\alpha ,b\beta \}} \mathcal{I}^{(\text{ii})}(z_0,\mu)
+C^{\,(\text{if})}_{\{a\alpha ,b\beta \}} \mathcal{I}^{(\text{if})}(z_0,\mu)
+C^{(\text{ff})}_{\{a\alpha ,b\beta \}} \mathcal{I}^{(\text{ff})}(z_0,\mu)
\right].
\end{equation}
The group theory factors are given by
\begin{align}
\label{eq:groupfactor}
C^{(\text{ii})}_{\{a\alpha ,b\beta \}}&= 2 \,
{\bf T}^{(r)a}_{a_1b_1}{\bf T}^{(r')a}_{a_2b_2} \delta_{\alpha \beta},\\
C^{(\text{if})}_{\{a\alpha ,b\beta \}}&= 
2 \left({\bf T}^{(r)a}_{a_1b_1}\delta_{a_2b_2}
  +\delta_{a_1b_1}{\bf T}^{(r')a}_{a_2 b_2}  \right)
{\bf T}^{(R_\alpha)a}_{\beta\alpha},\\
C^{(\text{ff})}_{\{a\alpha ,b\beta \}}&=\delta_{a_1b_1} \delta_{a_2b_2}
{\bf T}^{(R_\alpha)a}_{\beta\kappa}{\bf T}^{(R_\alpha)a}_{\kappa\alpha} 
=C_{R_\alpha}\delta_{a_1b_1} \delta_{a_2b_2}\delta_{\beta\alpha}.
\end{align}
For the initial-initial state diagrams it was used that diagrams with a 
soft-gluon coupling to two collinear particles in the same direction vanish.
The integrals are given by
\begin{align}
  \mathcal{I}^{(\text{ii})}(z_0,\mu) &=
\tilde\mu^{2\epsilon} (n\cdot \bar n)
\int_{-\infty}^0\!\!\!\!ds dt\, D^+(z_0 v +t n-s \bar n)
= -\frac{\Gamma(-\epsilon)}{8\pi^{2}}\frac{1}{\epsilon}
e^{\gamma_E\epsilon}\left(\frac{i z_0\mu}{2}\right)^{2\epsilon},\\
 \mathcal{I}^{\text{(if)}}(z^0,\mu)&=
\tilde\mu^{2\epsilon} (v\cdot n)
\int_{-\infty}^0\!\!\!\! ds dt\, D^+((z_0-s)v-n t)
=\frac{\Gamma(-\epsilon)}{16\pi^{2}}\frac{1}{\epsilon} e^{\gamma_E}
\left(\frac{i z_0\mu}{2}\right)^{2\epsilon},\\
 \mathcal{I}^{(\text{ff})}(z^0,\mu)&=
\tilde\mu^2 v^2
\int_{-\infty}^0\!\!\!\! ds dt\ D^+(v(z_0+ s-t))
=\frac{\Gamma(-\epsilon)}{8\pi^{2}} \frac{1}{(1-2\epsilon)}
e^{\gamma_E\epsilon}\left(\frac{iz_0\mu}{2}\right)^{2\epsilon}.
\label{eq:integralfactor}
\end{align}
Here $\tilde \mu^2=\mu^2
e^{\gamma_E}/(4 \pi)$ and an infinitesimal imaginary part $z_0\to z_0-i\delta$ 
is kept implicit.

It is useful to identify the amplitude $\mathcal{A}_{pp'\{a\}}$ with 
a vector in colour space denoted by 
$\ket{\mathcal{A}_{pp'}}$~\cite{Catani:1996vz,Kidonakis:1998nf}.  More
precisely the amplitude is a matrix element
$\mathcal{A}_{pp'\{a\}}=\braket{\{a\}|\mathcal{A}_{pp'}}$ with an
orthogonal basis $\ket{\{a\}}$.  In our notation an antiparticle in
representation $r$ is described as a particle in the representation $\bar
r$.  For incoming and outgoing particles the action of a generator acting on
particle $i$ in our conventions is given by
\begin{equation}
\begin{aligned}
\braket{\{b\}|{\bf T}_i^a|\mathcal{A}}&=
(-{\bf T}^{(r_i)a}_{a_ib_i})\mathcal{A}_{b_1 \dots a_i\dots b_n},
 &\text{incoming particle}\\
 \braket{\{b\}| {\bf T}_i^a|\mathcal{A}}&= 
 \mathcal{A}_{b_1 \dots a_i \dots b_n}{\bf T}^{(R_\alpha),a}_{b_ia_i},
& \text{outgoing particle}.
\end{aligned}
\end{equation}
Colour conservation implies the identity
\begin{equation}
\label{eq:conserve-colour}
\sum_i{\bf T}^a_i \ket{\mathcal{A}}=0.
\end{equation}
In this notation the decomposition~\eqref{eq:colour-amp} reads
\begin{equation}
\ket{\mathcal{A}_{pp'}}=\sum_i \ket{c^{(i)}} \mathcal{A}_{pp'}^{(i)},
\end{equation}
and the components of the soft function~\eqref{eq:soft-R} are expressed as
 $W^{R_\alpha}_{ii'}=\braket{c^{(i')}|\mathbf{W}^{R_\alpha}|c^{(i)}}$.  
Combining the results from eqs.~(\ref{eq:groupfactor}) to 
(\ref{eq:integralfactor}) and expanding in $\epsilon$, we find 
that the (unrenormalized) one-loop soft function 
in the colour-operator notation reads:
\begin{eqnarray}
\hat {\bf W}^{(1)\,R_\alpha}(L) &=& -\left(
({\bf T}_1+{\bf T}_2)\cdot {\bf T}_3+2 {\bf T}_1\cdot{\bf T}_2\right)
\left(\frac{2}{\epsilon^2}+\frac{2}{\epsilon} L
  + L^2+\frac{\pi^2}{6}\right)
+{\bf T}_3^2\left(\frac{2}{\epsilon}+2 L+4\right)
\nonumber\\
&=&\left({\bf T}^2_1+{\bf T}^2_2\right)
\left(\frac{2}{\epsilon^2}+\frac{2}{\epsilon} L
  + L^2+\frac{\pi^2}{6}\right)
+2\,{\bf T}_3^2\left(\frac{1}{\epsilon}+ L+2\right).
\end{eqnarray}
Here we have introduced the variable~\cite{Korchemsky:1992xv}
\begin{equation}
\label{eq:def-L}
L=2\ln\left(\frac{i z_0\mu e^{\gamma_E}}{2}\right),
\end{equation}
and used colour conservation~\eqref{eq:conserve-colour} to arrive 
at the second equality. 

The components $W_{ii'}^{R_\alpha}$ entering the factorization
formula~\eqref{eq:fact} can be obtained from the above result by contracting
with the elements of the colour basis $c_{\{a\alpha\}}^{(i)R_\alpha}$
according to~\eqref{eq:soft-coulomb-2}.  As discussed in
section~\ref{sec:colour-soft}, a large number of matrix elements are
zero by construction and the soft matrices assume a block-diagonal
form (c.f.~\eqref{eq:soft-block}).  Since the tree-level and one-loop
soft functions are proportional to the unit matrix in colour space and
the colour tensors in our basis are given in terms of the
Clebsch-Gordan coefficients~\eqref{eq:c-ralpha}, we find that the
components~\eqref{eq:soft-coulomb-2} are diagonal due to the orthogonality
of the Clebsch-Gordan coefficients:
\begin{equation}
 \hat W^{R_\alpha}_{ii'}(L,\mu)= \hat W^{R_\alpha}_{i}(L,\mu)\;\delta_{ii'}\;
 \delta_{R_\alpha R_\beta}.
\end{equation}
In agreement with the general results of section~\ref{sec:colour-soft} these
elements are non-vanishing only if the final state representation
$R_\alpha$ is identical to that in the pair $P_i=(r_\alpha,R_{\beta})$
that defines the basis element $c^{(i)}$.
The diagonal elements 
at tree- and one-loop level are given by
\begin{align}
 \hat W^{(0)R_\alpha}_{i}(L,\mu)&=1, \\
 \hat W^{(1)R_\alpha}_{i}(L,\mu)&=\left(
C_r+C_{r'}\right)\left(\frac{2}{\epsilon^2}+\frac{2}{\epsilon} L
  + L^2+\frac{\pi^2}{6}\right)
+2C_{R_\alpha}\left(\frac{1}{\epsilon}+ L+2\right) .
\label{eq:soft-one}
\end{align}
As shown in appendix~\ref{app:fourier} the Fourier transform of this result
agrees with~\cite{Idilbi:2009cc} for the special case of the production of a
colour-octet particle from gluon fusion.

\subsection{Renormalization group equations and anomalous dimensions}
\label{sec:gamma}

In the following we will provide the evolution equations of the hard function
$H_{ii'}$ and the soft function $W^{R_\alpha}_{ii'}$ and determine the
relevant anomalous dimensions at the two-loop level, as required for
resummation at NNLL accuracy.
 As a result of the rewriting of the soft function
in section~\ref{sec:colour}, the soft anomalous-dimension matrix for pair
production at threshold is identical to that of a two-to-one scattering
process with two massless legs and one massive leg.  Employing results from a
recent analysis of constraints from soft-collinear factorization on the
structure of infrared~(IR) singularities of scattering amplitudes with massive
particles~\cite{Becher:2009kw}, we extract an analytical expression for the
two-loop soft anomalous dimension at threshold.  Explicit expressions for the
resummed cross section in momentum space will be given in~\cite{BFS}, but we
stress that the results given here are also applicable to the resummation in
Mellin-moment space. The precise relation to the formalism in Mellin space is
discussed in section~\ref{sec:compare}.

As mentioned above, in the effective field theory treatment
of the factorization formula the hard function is defined in terms of 
short-distance coefficients $C_{pp'}^{(i)}$ that are obtained from 
the components $\mathcal{A}^{(i)}_{pp'}$ of the scattering
amplitude~\eqref{eq:colour-amp} evaluated  at threshold~\cite{BFS}.
After renormalization of the ultraviolet
divergences, the short-distance coefficients contain further IR divergences
that match the ultraviolet divergences of the long-distance objects 
in the factorization formula.
The IR-renormalized coefficients obtained by minimal subtraction of 
the IR-poles satisfy an evolution equation of the form
\begin{equation}
\label{eq:RGE-C}
\frac{d}{d\ln\mu} C_{pp'}^{(i)}(M,\mu)=\Gamma_{ij}(M,\mu) 
\,C_{pp'}^{(j)}(M,\mu)
\end{equation}
with an
anomalous-dimension matrix whose form is constrained by soft and collinear
factorization~\cite{Becher:2009cu,Gardi:2009qi,Becher:2009qa,Becher:2009kw}.
Since we have shown that the (leading) 
soft function is diagonal for all cases relevant
to hadron-collider processes, only the diagonal elements of the hard
function, $H_{ii}\equiv H_{i}$ enter the formula for the production cross
section~\eqref{eq:fact}. They satisfy the evolution equation
\begin{equation}
\label{eq:rge-hard}
\frac{d}{d\ln\mu}H_{i}(M,\mu)=2 \,\text{Re}\,\Gamma_i(M,\mu) \,H_{i}(M,\mu)
\end{equation}
with  $\Gamma_{ii}\equiv \Gamma_{i}$.
As shown in appendix~\ref{app:gamma} the results of~\cite{Becher:2009kw}
constrain the anomalous dimension to be of the form
\begin{equation}
\label{eq:gamma-sc}
\Gamma_i(M,\mu)=\frac{1}{2}\,\gamma_{\text{cusp}}\left[
(C_r+C_{r'})\left(\ln\left(\frac{4M^2}{\mu^2}\right)-i \pi\right)
+i\pi C_{R_\alpha}\right]+\gamma^V_i.
\end{equation}
Here we have introduced the coefficient $\gamma_{\text{cusp}}$ by
writing the cusp anomalous dimension for a massless parton in the
representation $r$ in the form $\Gamma_{\text{cusp}}^r=C_r
\gamma_{\text{cusp}}$ consistent with Casimir scaling which is
appropriate at least up to three-loop order. The explicit one- and 
two-loop results for
all anomalous dimensions needed in this section are collected in
appendix~\ref{app:gamma}. The coefficient
$\gamma_{\text{cusp}}$ is known to three-loop
order~\cite{Moch:2004pa}, and the $\alpha_s (\alpha_s n_f)^k$ 
terms are known to all orders \cite{Beneke:1995pq}.

Adopting the result of~\cite{Becher:2009kw} for the structure of the anomalous
dimension matrix,  at least up to the two-loop level the anomalous dimension
$\gamma^V_i$ can be written in terms of single-particle anomalous dimensions:
\begin{equation}
\label{eq:gamma-v}
\gamma^V_i=\gamma^r+\gamma^{r'}+\gamma_{H,s}^{R_\alpha}.
\end{equation}
The one- and two-loop anomalous-dimension coefficients $\gamma^r$ of
massless quarks, $\gamma^q=\gamma^3$, and gluons, $\gamma^g=\gamma^8$
are given in appendix A of ref.~\cite{Becher:2009qa}.
The anomalous dimension $\gamma_{H,s}^{R_\alpha}$ is related to a massive
particle in the final state representation $R_\alpha$ in the pair
$P_i=(r_\alpha',R_\alpha)$ defining the colour basis element
$c^{(i)}$ with index $i$. 

It should be mentioned that eqs.~(\ref{eq:gamma-sc}) and (\ref{eq:gamma-v})
are derived from ref.~\cite{Becher:2009kw}, where it is assumed that 
the two heavy particles have fixed but unequal velocities, when the 
poles in $\epsilon$ of the hard amplitude are extracted. This is 
different from the limit we consider here, where $\beta\to 0$ 
before the limit $\epsilon\to 0$ and before the loop integrations 
are performed, which corresponds to the 
threshold expansion as defined in~\cite{Beneke:1997zp}.  The order of 
limits does not commute, and by expanding in $\beta$ first new IR 
divergences appear in the hard region that do not
correspond to UV divergences in the soft and collinear but  
in the potential region of the threshold expansion. 
The complete result for the scale dependence (\ref{eq:rge-hard}) 
of the hard coefficient may thus contain additional terms related 
to the ultraviolet divergences of the higher-dimensional heavy-quark 
potentials and soft functions. In deriving eq.~(\ref{eq:gamma-sc}) 
from the $2\to 1$ process with a single particle in
representation $R_\alpha$ we implicitly set to zero the scale
dependence of the hard function related to the contribution 
from potential divergences and the higher-dimensional soft functions. 
This allows us to 
relate the anomalous dimensions (\ref{eq:gamma-sc}) and
(\ref{eq:gamma-v}) directly to the one of the leading soft 
function~$W_{i i'}^{R_\alpha}$.

For the case of a heavy quark an analytical result for the two-loop anomalous
dimension~$\gamma^Q=\gamma_{H,s}^{3}$ has been extracted 
in~\cite{Becher:2009kw} from the anomalous dimension of the heavy-light quark
current in SCET.  In order to generalize this result to arbitrary
representations $R$, we observe, following~\cite{Becher:2009kw}, that
$\gamma_{H,s}^{R}$ appears in the anomalous dimension of the HQET heavy-heavy
current for a heavy particle in representation $R$,
\begin{equation}
\label{eq:heavy}
  \Gamma_{J_{hh}^R}=C_R\,\gamma_{\text{cusp}}(\beta,\alpha_s)+2\gamma^R_{H,s}.
\end{equation}
The cusp anomalous dimension $\gamma_{\text{cusp}}(\beta,\alpha)$ is a
function of the cusp angle $\cosh\beta=v_1\cdot v_2$, with $v_{1,2}$ the 
four-velocities of the heavy particles. The anomalous dimension 
$\Gamma_{J_{hh}}$
for heavy quarks is available at two-loop order~\cite{Korchemsky:1991zp}.  For
large cusp angle the massive cusp anomalous dimension is related to the cusp
anomalous dimension for massless particles according to~\cite{Becher:2009kw}
\begin{equation}
\label{eq:cusp-infty}
 \gamma_{\text{cusp}}(\beta,\alpha_s)\rightarrow
\gamma_{\text{cusp}}(\alpha_s)\,\beta +\dots\;,
\end{equation}
where the remainder vanishes for $\beta\to \infty$. It follows that 
the heavy particle soft anomalous dimension
$\gamma^R_{H,s}$ can be obtained as one-half of the constant coefficient in
the anomalous dimension of the heavy-heavy current~\eqref{eq:heavy} in the
limit where the cusp angle goes to infinity.  Since the anomalous dimension of
the heavy-heavy formfactor in HQET is related to the expectation value of a
Wilson line~\cite{Korchemsky:1991zp} the colour structure is constrained by
the non-abelian exponentiation theorem~\cite{Gatheral:1983cz,Frenkel:1984pz}.
For a heavy particle in the representation $R$, at the two-loop level only the
colour structures $C_R C_A$ and $C_R T_F n_f$ appear.  We therefore obtain the
one- and two-loop anomalous dimensions for a heavy particle in an arbitrary
representation from the result for a heavy quark by simple Casimir scaling,
\begin{equation}
\label{eq:gamma-H-casimir}
\gamma_{H,s}^{R_\alpha}=C_{R_\alpha}\gamma_{H,s},
\end{equation}
with $\gamma_{H,s}=\gamma^{Q}/C_F$. Adopting the two-loop
anomalous dimension of the HQET formfactor~\cite{Korchemsky:1991zp} in the
explicit formulation in terms of polylogarithms given
in~\cite{Kidonakis:2009ev} we obtain
\begin{equation}
\label{eq:gamma-H}
\begin{aligned}
\gamma_{H,s}^{(0)}&=-2,\\
\gamma_{H,s}^{(1)}&=-C_A\left(\frac{98}{9}-\frac{2\pi^2}{3}+4\zeta_3\right)
+\frac{40}{9} T_F n_f,
\end{aligned}
\end{equation} 
where the loop expansion of the anomalous dimensions is defined as
in~\eqref{eq:gamma-alpha}.  The one-loop expression
in~\eqref{eq:gamma-H} agrees with the well-known result that the
one-loop soft anomalous dimension is proportional to the quadratic
Casimir of the final state
system~\cite{Bonciani:1998vc,Kidonakis:1997gm,Kulesza:2008jb}. The
two-loop expression for the anomalous dimension~\eqref{eq:gamma-v} for 
heavy-particle pair production in an arbitrary colour representation is a new
result.  The Casimir scaling of the single-particle anomalous
dimension~\eqref{eq:gamma-H-casimir} was also noted in the published version
of~\cite{Becher:2009kw}. 

The factorization scale independence of the hadronic cross section in the
threshold region can be used to obtain the
evolution equation of the soft function as~\cite{BFS}
\begin{equation}
\label{eq:rge-soft}
\frac{d}{d\ln\mu}  \hat W^{R_\alpha}_i(L)
  =  \left(
    (\Gamma_{\text{cusp}}^{r}+\Gamma_{\text{cusp}}^{r'})L 
    -2\gamma_{W,i}^{R_\alpha}\right) \hat W^{R_\alpha}_i(L),
\end{equation}
where the anomalous dimension of the soft function is obtained from the
anomalous dimension of the hard function by 
adding anomalous dimensions $\gamma^{\phi,r}$ entering the
evolution equations of the parton distribution function for a parton
 in the representation $r$ in the $x\to 1$ limit (see~\ref{eq:def-gamma-phi}):
\begin{equation}
\label{eq:relation-soft-gamma}
\gamma_{W,i}^{R_\alpha}=\gamma^V_i+\gamma^{\phi,r}+\gamma^{\phi,r'}.
\end{equation}
The anomalous dimensions $\gamma^{\phi,r}$ for 
 quarks and gluons are available up to three-loop
order~\cite{Moch:2004pa}. The one- and two-loop results are collected
in appendix~\ref{app:gamma}. We note that 
eq.~(\ref{eq:relation-soft-gamma}) is derived under the assumption 
that the Coulomb function $J_{R_\alpha}$ is scale-independent, 
$d J_{R_\alpha}/d\ln\mu=0$, which is no longer true starting from 
NNLL in the non-relativistic expansion. However, as remarked above 
this type of scale dependence is related to terms that we consistently
dropped in the derivation of eqs.~(\ref{eq:gamma-sc}) and 
(\ref{eq:gamma-v}), so the two-loop soft 
anomalous dimension of the Wilson line considered here is unaffected 
by these complications.

Analogously to the anomalous dimension $\gamma^V_i$ of the hard
function~\eqref{eq:gamma-v}, at least up to the two-loop level the anomalous
dimension of the soft function~\eqref{eq:relation-soft-gamma} can be written
in terms of separate single-particle contributions
\begin{equation}
\label{eq:gamma-W}
\gamma_{W,i}^{R_\alpha}=\gamma_{H,s}^{R_\alpha}+\gamma^r_s+\gamma_s^{r'}
\end{equation}
with 
\begin{equation}
\label{eq:gamma-s-massless}
 \gamma^r_s=\gamma^r+\gamma^{\phi,r}.
\end{equation}
Up to the two-loop level, the anomalous dimensions for quarks
$\gamma^q_s=\gamma^3_s$ and gluons $\gamma^g_s=\gamma^8_s$ are related
by Casimir scaling,
\begin{equation}
  \gamma^r_s=C_r\gamma_s ,
\end{equation}
where the one- and
two-loop coefficients are~\cite{Becher:2007ty}:
\begin{equation}
\label{eq:gamma-s}
\begin{aligned}
 \gamma_s^{(0)}&=0 ,\\
  \gamma_s^{(1)}&=C_A
  \left(-\frac{404}{27}+\frac{11\pi^2}{18}+14\zeta_3\right)
  +T_F n_f\left(\frac{112}{27}-\frac{2\pi^2}{9}\right).
\end{aligned}
\end{equation}

The evolution
equations~\eqref{eq:rge-hard} and~\eqref{eq:rge-soft} generalize the
corresponding equations for the Drell-Yan process~\cite{Becher:2007ty}
and Higgs production~\cite{Ahrens:2008nc} to processes with a heavy
particle pair in the final state. The evolution equation for the soft
function in momentum space involves distributions and can be solved in
Mellin-moment space~\cite{Korchemsky:1992xv} or directly in momentum
space~\cite{Becher:2006nr,Becher:2006mr,Becher:2007ty} using a Laplace
transform. Using the two-loop result~\eqref{eq:gamma-H}, the
anomalous dimension in the evolution equation~\eqref{eq:rge-soft} 
is known with the accuracy required for NNLL resummation.

\subsection{Relation to the formalism in Mellin-moment space}

\label{sec:compare}

In the applications of resummation in
Mellin space~\cite{Catani:1996yz,Bonciani:1998vc,Moch:2008qy,Cacciari:2008zb,Kidonakis:2008mu,Kulesza:2008jb,Kulesza:2009kq}
  the Mellin moments of the partonic cross section with respect to the
  variable $\rho=4M^2/\hat s$ are written in the form
\begin{equation}
\label{eq:sigma-mellin}
\begin{aligned}
\hat \sigma^{N}_{pp',R_\alpha}(M^2,\mu)&\equiv
\int_0^1d\rho \rho^{N-1} \hat \sigma_{pp',R_\alpha}(4M^2/\rho,\mu)\\
&=\hat \sigma^{(0)N}_{pp',R_\alpha}(M^2,\mu) g^0_{pp,R_\alpha}(M^2,\mu)\;
\exp\left( G_{pp',R_\alpha}^{N+1}(M^2,\mu^2)\right).
\end{aligned}
\end{equation}
Here $ \hat \sigma_{pp',R_\alpha}$ is the partonic cross section for
the production of a heavy-particle final state pair in the
representation $R_\alpha$, $\sigma^{(0)N}_{pp',R_\alpha}$ are the
Mellin moments of the Born cross section, the matching functions
$g^0_{pp,R_\alpha}$ collect the $N$-independent corrections, and the
exponent $G_{pp',R_\alpha}^{N}$ has the form
\begin{eqnarray}
G_{pp',R_\alpha}^{N}(M^2,\mu^2)&=&
\int_0^1 dz \,\frac{z^{N-1}-1}{1-z}\Biggl[\,
\int_{\mu^2}^{4M^2(1-z)^2} \frac{dq^2}{q^2} 
\left((A_p\left(\alpha_s(q^2)\right)+A_{p'}\left(\alpha_s(q^2)\right)\right)
\nonumber \\
&& \hspace*{0cm} 
+ \, D^{R_\alpha}_{pp'\to HH'}\left(\alpha_s(4M^2(1-z)^2)\right)\Biggr].
\label{eq:mellin}
\end{eqnarray}
The coefficients $A_{p}$ contain the effect of collinear radiation off
the incoming partons and are identical to the cusp anomalous
dimension $\Gamma_{\text{cusp}}^r$. The coefficient $D$ describes soft
radiation and can be written as a sum of terms related to the
incoming partons and the final state system
\begin{equation}
\label{eq:D}
D^{R_\alpha}_{pp'\to HH'}(\alpha_s)=
\frac{1}{2}\,\big(D_p(\alpha_s)+D_{p'}(\alpha_s)\big)
  +D^{R_\alpha}_{HH'}(\alpha_s).
\end{equation}
Applications of resummation in Mellin space often use an
equivalent form of~\eqref{eq:sigma-mellin} where $N$-independent terms
contained in the large-$N$ expansion of~\eqref{eq:mellin} are not included in
the exponent. This implies a redefinition of the matching functions
$g^0_{pp,R_\alpha}(M^2,\mu)$ but no change in the $A$ and $D$ 
coefficients.

The coefficient $D^{R_\alpha}_{pp'\to HH'}$ is not identical to the
anomalous dimension of the soft function $\gamma^{R_\alpha}_W$ because
the Mellin transform of the resummed cross section in momentum space
involves the fixed-order soft function at the scale $\mu_s\sim
M/N$~\cite{Korchemsky:1993uz,Becher:2007ty} while in the Mellin-space
formula~\eqref{eq:sigma-mellin} all the $N$-dependent soft corrections
are exponentiated. Furthermore the form of the exponent is different
in the two approaches.  The relation of different forms of
exponentiated expressions in Mellin space is discussed
in~\cite{Catani:2003zt,Idilbi:2006dg,Becher:2006mr,Becher:2007ty,Beneke:1995pq}.
But for heavy-particle pair production near threshold, the structure
of the resummed expressions in Mellin space and momentum space is
identical to that for the Drell-Yan process, where the relation
between the two formalisms was obtained in eq.~(71)
of~\cite{Becher:2007ty}.  This then implies a relation of the
coefficient~\eqref{eq:D} to the anomalous dimension of the soft
function~\eqref{eq:gamma-W} given by
\begin{eqnarray}
  e^{2\gamma_E\nabla}\,\Gamma(1+2\nabla)D^{R_\alpha}_{pp'\to HH'}
  &=&2\,\gamma_{W,i}^{R_\alpha}+2\nabla \ln
  \tilde{s}^{R_\alpha}_{i}(0,\mu)
  \nonumber \\[0.2cm]
  && -\, \frac{e^{2\gamma_E\nabla}\Gamma(1+2\nabla)-1}{\nabla}
  (\Gamma_{\text{cusp}}^r+\Gamma_{\text{cusp}}^{r'}),
\label{eq:D-gamma}
\end{eqnarray}
where
\begin{equation}
  \nabla=\frac{d}{d\ln\mu^2}
=\frac{\beta(\alpha_s)}{2}\frac{\partial}{\partial\alpha_s}.
\end{equation}
We also introduced the 
Laplace-transform of the soft function~\cite{Becher:2006nr,Becher:2006mr} 
with respect to the variable $s =
1/(e^{\gamma_E} \mu \,e^{\rho/2})$
\begin{equation}
\tilde{s}^{R_\alpha}_{i}(\rho,\mu) =\int_{0}^{\infty} d \omega \,e^{-s \omega}
\, \overline W_{i}^{R_\alpha}(\omega,\mu), 
\end{equation}
where we have defined the $\overline{\text{MS}}$-renormalized soft
function $\overline W^{R_\alpha}_i$.  Since the soft function in
position space $\hat W(z_0,\mu)$ depends on the arguments solely
through the variable $(i z_0\mu e^{\gamma_E}/2)=e^{L/2}$ with $L$
defined in~\eqref{eq:def-L}, it is easy to see that the function
$\tilde s(\rho)$ is obtained by simply replacing $L\to -\rho$ in the
$\overline{\text{MS}}$-renormalized result for the soft function in
position space.
Expanding  the relation~\eqref{eq:D-gamma} counting $\nabla\sim \alpha_s$
we obtain the terms relevant to determine $D$ at the two-loop level:
\begin{equation}
 D^{R_\alpha}_{pp'\to HH'}
  =2\gamma_{W,i}^{R_\alpha}+2\nabla \ln \tilde s^{R_\alpha}_i(0,\mu)
  -\frac{\pi^2}{3}\nabla
  (\Gamma_{\text{cusp}}^r+\Gamma_{\text{cusp}}^{r'})+\mathcal{O}(\alpha_s^3).
\end{equation}

Using the loop expansion of the soft function~\eqref{eq:soft-alpha},
the anomalous dimensions~\eqref{eq:gamma-alpha} and the beta function, 
\begin{equation}
  \beta(\alpha_s)=\mu\frac{\partial \alpha_s}{\partial \mu}
=-2\alpha_s\sum_{n=0}^\infty 
 \left(\frac{\alpha_s(\mu)}{4\pi}\right)^{n+1}\beta_n
\end{equation}
with $\beta_0=11/3 \,C_A-4/3 \,T_F n_f$,
we find up to the two-loop level
\begin{align}
  D^{(0)R_\alpha}_{pp'\to HH'}&= 2\,\gamma_{W,i}^{(0)R_\alpha},\\
 D^{(1)R_\alpha}_{pp'\to HH'}&=
2\,\gamma_{W,i}^{(1)R_\alpha}-2\beta_0\left(\tilde s^{(1)R_\alpha}_i(0,\mu)
-\frac{\pi^2}{6}(\Gamma_{\text{cusp}}^{(0)r}+\Gamma_{\text{cusp}}^{(0)r'}) \right)
\nonumber\\
&= 2\,\gamma_{W,i}^{(1)R_\alpha}
+\beta_0\left(\pi^2(C_r+C_{r'})-8C_{R_\alpha}\right),
\label{eq:D-2}
\end{align}
where we have used $\Gamma_{\text{cusp}}^{(0)r}=4C_r$ and the one-loop
soft function~\eqref{eq:soft-one} in the last step (with $L=0$ and 
the $1/\epsilon$ poles discarded due to the 
$\overline{\rm MS}$ subtraction).

Using the decomposition~\eqref{eq:gamma-W} of the anomalous dimension
of the soft function and applying~\eqref{eq:D-2} to Drell-Yan and 
Higgs production, in which case $r=r'$, $C_{R_\alpha}=0$ (colour
singlet final state) and $\gamma_{W,i}^1=2 \gamma_s^r$ due to 
(\ref{eq:gamma-W}), we obtain the coefficients $D_q$ and $D_g$ introduced
in~\eqref{eq:D}:
\begin{equation}
\label{eq:D-light}
  D_p=(4\gamma_s^{(1)r}+2\pi^2 C_r\beta_0)\left(\frac{\alpha_s}{4\pi}\right)^2+
\mathcal{O}(\alpha_s^3).
\end{equation}
The result for $D_q$ agrees with eq.~(72) of~\cite{Becher:2007ty} and
the explicit expression obtained using~\eqref{eq:gamma-s} agrees with the one
used in eq.~(A.3) of~\cite{Moch:2008qy}.

Turning to the soft anomalous dimension related to the heavy particle pair, 
we find that the relation of the coefficients appearing in the 
Mellin-space approach to the soft anomalous dimension obtained 
in~\eqref{eq:gamma-H} is given by 
\begin{align}
  D^{(0)R_\alpha}_{HH'}&=2 \gamma_{H,s}^{(0)R_\alpha}, \\
   D^{(1)R_\alpha}_{HH'}&=2\gamma_{H,s}^{(1)R_\alpha}-8\beta_0C_{R_\alpha}.
\end{align}
The one-loop coefficient $ D_{HH'}^{(0)R_\alpha}=-4C_{R_\alpha}$ is in
agreement with previous results for top-quark and gluino
production~\cite{Bonciani:1998vc,Kidonakis:1997gm,Kulesza:2008jb}.  The result
for the two-loop coefficient is new and shows
that at NNLL level there is a non-trivial relation between the soft anomalous
dimension and the resummation coefficients if eq.~\eqref{eq:sigma-mellin}
is used for the resummed cross section.
 Our result for the two-loop soft anomalous
 dimension~\eqref{eq:gamma-H} allows to obtain for the first time the
 two-loop coefficient for soft radiation off a massive particle pair
 at threshold for all possible colour states of the heavy-particle
 system:
\begin{equation}
\label{eq:D-H}
 D_{HH'}^{(1)R_\alpha}=-C_{R_\alpha}C_A\left(\frac{460}{9}-
\frac{4\pi^2}{3}+8\zeta_3\right)
 +\frac{176}{9}C_{R_\alpha} T_F n_f.
\end{equation}
  Our result~\eqref{eq:D-H} differs from the result quoted
  in~\cite{Moch:2008qy} for the special case of a colour-octet final
  state by a term $8C_{A} \left[C_A(1-\zeta_3)-\beta_0\right]$.
  However, the result in~\cite{Moch:2008qy} was obtained from an analysis of
  the singularities of the two-loop massive quark
  form factor~\cite{Bernreuther:2004ih,Mitov:2006xs} in the limit of
  light quark masses using the unjustified --  and as it turns out
  incorrect -- assumption that the factor of
  proportionality of the two-loop and one-loop soft anomalous
  dimension matrices in the massless limit is identical to the one at 
threshold. 

\section{Conclusions and outlook}
We have performed a detailed study of the soft function relevant to
threshold resummation of production processes of heavy coloured
particle pairs at hadron colliders. We have given a precise
formulation of the physical picture of soft-gluon radiation coupling
to the total colour charge of the heavy-particle pair. This has
allowed us to construct a colour basis that diagonalizes the soft
function to all orders in perturbation theory.  Explicit expressions
for all production processes of top quarks, squarks and gluinos have been
provided.  We have calculated the one-loop soft function for arbitrary
colour representations of initial and final state particles and used
recent new insights into soft-collinear factorization to obtain the
two-loop soft anomalous dimension. This supplies the
process-independent ingredients for NNLL resummation of threshold logarithms
in arbitrary production processes of heavy coloured particles at
hadron colliders. A complete NNLL resummation in the sense of 
eq.~(\ref{eq:syst}) further needs the colour-separated one-loop 
short-distance coefficients, as well as the summation of logarithms 
associated with subleading terms in $\beta$ in the non-relativistic 
expansion. 

In a subsequent publication~\cite{BFS} we will give a derivation 
of the factorization formula~\eqref{eq:fact} that demonstrates the 
factorization of Coulomb gluon exchange from soft-gluon radiation 
using field redefinitions in an effective
field theory. The formula thus provides a theoretically 
clearly defined separation of hard,
Coulomb and soft effects and can be used for a combined resummation of
soft and Coulomb gluons. This will be discussed for the case 
of squark-antisquark production, where the
effect of Coulomb resummation may be of similar order as
the effect of soft gluon radiation~\cite{Kulesza:2009kq}.

\subsection*{Acknowledgements}
We thank Thomas Becher, Lance Dixon, Sven Moch and
Matthias Neubert for useful discussions.  M.B. thanks the CERN theory
group for its hospitality, while part of this work was done. The work
of M.B. is supported in part by the DFG
Sonder\-for\-schungs\-bereich/Trans\-regio~9 ``Computergest\"utzte
Theoretische Teilchenphysik''.

\vspace*{0.2cm}\noindent
After completion of this work we learnt of an independent 
calculation~\cite{Czakon:2009zw}
of the two-loop soft anomalous dimension (\ref{eq:D-H}) for the case
of heavy-quark production ($C_{R_\alpha}=C_A$) by a different method, 
which is in agreement with our result. We thank Michal Czakon for 
comparing results prior to publication.

\section*{Note added}

During the review process of this paper refs.~\cite{Ferroglia:2009ep}
appeared, in which the authors reported a non-vanishing result for the
three-particle contributions to the two-loop soft anomalous dimension
of amplitudes with two massive particles and any number of massless
particles near threshold, which is also not diagonal in the colour
basis discussed in the present paper. We emphasize that these findings
are not in contradiction with the results reported here. The authors
of refs.~\cite{Ferroglia:2009ep} calculate the $1/\epsilon$ poles away
from threshold and then take the limit $\beta\to 0$, resulting in
logarithms of $\beta$ in the anomalous dimension and non-zero
three-particle correlations.  In the present approach, where the
expansion in $\beta$ is constructed directly within the
non-relativistic effective field theory framework, the extra
logarithms are related to the potential region and the three-particle
correlations to higher-dimensional soft functions, both belonging to
the possible NNLL terms mentioned in the text that arise from
subleading heavy-quark potentials, and from $O(\beta)$ terms
interfering with Coulomb singularities. Thus, the derivation of the
diagonal colour basis and two-loop anomalous dimension for the soft
function with two equal heavy particle velocities discussed in the
present paper remains valid in the light of the results of
refs.~\cite{Ferroglia:2009ep}, but should not be expected to apply to
higher-dimensional soft functions. (In fact, the emission of a soft
gluon due to the subleading $\vec{x}\cdot \vec{E}$ interaction 
mentioned in section~\ref{sec:nnll} leads
to a change in the colour state of the heavy-particle pair, and the
corresponding soft function must therefore be off-diagonal.) Finally,
let us mention that while the results of refs.~\cite{Ferroglia:2009ep}
may be used to determine all logarithmic terms at $O(\alpha_s^2)$, the
summation of NNLL logarithms in higher orders in the strong coupling
expansion requires an analysis of soft and potential divergences
directly at threshold.

%%%%%%%%%%%%%%%%%%%%%%%%%%%%%%%%%%%%%%%%%%%%%%%%%%%%%%%%%%%%%%
\appendix

\section{Technical details on the colour structure }
\label{app:colour-details}

\subsection{Construction of the colour basis}
To see that the basis can always be chosen as in~\eqref{eq:prod-basis},
we use the completeness relation~\eqref{eq:unitary-clebsch} of the 
Clebsch-Gordan coefficients to decompose the scattering
amplitude~\eqref{eq:colour-amp} according to
\begin{equation}
\mathcal{A}_{\{a\}}=
\sum_{r_\alpha,R_\beta} \mathcal{A}_{\alpha\beta}\,
C^{r_\alpha}_{\alpha a_1a_2} C^{R_{\beta}\ast}_{\beta a_3a_4}
\end{equation}
with coefficients
\begin{equation}
\label{eq:A-aux}
\mathcal{A}_{\alpha\beta}=C^{r_\alpha\ast}_{\alpha b_1b_2}
C^{R_{\beta}}_{\beta b_3b_4}\mathcal{A}_{\{b\}}.
\end{equation}
The fact that the amplitude $\mathcal{A}_{\{a\}}$ is colour
conserving, i.e.~satisfies an identity analogous
to~\eqref{eq:global-basis}, and the invariance condition of the
Clebsch-Gordan coefficients imply that the coefficients~\eqref{eq:A-aux} are
invariant tensors under transformations in $r_\alpha\otimes
\overline{R}_\beta$:
\begin{equation}
\mathcal{A}_{\alpha \beta}= U^{(R_\beta)\dagger}_{\beta\delta}
\mathcal{A}_{\gamma\delta}\,
U^{(r_\alpha)}_{\gamma\alpha} .
\end{equation}
Hence according to Schur's Lemma $\mathcal{A}_{\alpha \beta}$ is non-vanishing
only if $r_\alpha$ and $R_\beta$ are
equivalent irreducible representations, i.e.~belong to one of the
pairs $P_i=(r_\alpha,R_\beta)$ of equivalent representations
appearing in the decompositions of the initial and final state system
into irreducible representations introduced 
above eq.~\eqref{eq:tensor-88}. In this case $\mathcal{A}_{\alpha\beta}=
\tilde {\mathcal{A}}^{(i)} \,\delta_{\alpha\beta}$ for some coefficient
  $\tilde {\mathcal{A}}^{(i)}$.  We have therefore derived the 
decomposition
\begin{equation}
\label{eq:colour-amp-2}
\mathcal{A}_{\{a\}}=
\sum_{P_i} \tilde {\mathcal{A}}^{(i)} \,
C^{r_\alpha}_{\alpha a_1a_2} C^{R_{\alpha}\ast}_{\alpha a_3a_4}
\end{equation}
 of the amplitude. Using the orthogonality of the Clebsch-Gordan 
coefficients we find
 that the constant $\tilde {\mathcal{A}}^{(i)}$ is, up to
 normalization, given by the components of the amplitude in the
 basis~\eqref{eq:prod-basis}
\begin{equation}
  \tilde {\mathcal{A}}^{(i)}=\frac{1}{\text{dim}(r_\alpha)}\,
    C^{r_\alpha\ast}_{\alpha a_1a_2} C^{R_{\alpha}}_{\alpha a_3a_4} 
    \mathcal{A}_{\{\alpha\}}
    \equiv \frac{1}{\sqrt{\text{dim}(r_\alpha)}} \,\mathcal{A}^{(i)}.
\end{equation}
 The identity~\eqref{eq:colour-amp-2} is therefore
precisely the decomposition of the amplitude into the
basis~\eqref{eq:prod-basis} for the coefficients.

\subsection{Identities for Wilson lines}
In order to rewrite the soft function in terms of the Wilson line for a 
single heavy particle in a representation $R_\alpha$ we used the
identity~\eqref{eq:combine-wilson} and its complex conjugate
\begin{equation}
S^{(R')\dagger }_{v,b_2a_2}S^{(R)\dagger }_{v,b_1a_1} 
C^{R_\alpha\ast}_{\alpha a_1a_2}
=C^{R_\alpha\ast}_{\beta b_1b_2}S^{(R_\alpha)\dagger}_{v,\beta\alpha}.
\label{eq:combine-wilson-ad}
\end{equation}

To prove identity~\eqref{eq:combine-wilson} we observe that the 
Wilson line~\eqref{eq:def-wilson} solves the differential equation
\begin{equation}
\label{eq:diff-wilson}
(v\cdot D) \,S^{(R)}_v(x_0)
=(v\cdot \partial -i g_s v\cdot A^a {\bf T}^{a(R)})S^{(R)}_v(x_0)=0.
\end{equation}
Then using the relation 
\begin{equation}
C^{R_\alpha}_{\alpha a_1a_2}\left({\bf T}^{a(R)}_{a_1b_1} \delta_{a_2b_2}
+\delta_{a_1b_1} {\bf T}^{a(R')}_{a_2b_2}\right)=
{\bf T}^{a(R_\alpha)}_{\alpha\beta}
C^{R_\alpha}_{\beta b_1b_2},
\end{equation}
obtained from the invariance condition~\eqref{eq:clebsch-transform} for 
infinitesimal transformations, we see that the left-hand side
of~\eqref{eq:combine-wilson} also satisfies~\eqref{eq:diff-wilson} for a 
Wilson line in the representation $R_\alpha$, if the Wilson lines 
$S^{(R)}_{v}$ and $S^{(R')}_{v}$ satisfy the analogous definitions 
in their representations:
\begin{eqnarray}
&& (v\cdot D)_{\alpha\beta}C^{R_\alpha}_{\beta a_1a_2}S^{(R)}_{v,a_1b_1}
S^{(R')}_{v,a_2b_2}
\nonumber\\[0.2cm]
&& \hspace*{1cm} = \, C^{R_\alpha}_{\alpha a_1a_2}\left[
((v\cdot D)S^{(R)}_{v})_{a_1b_1}S^{(R')}_{v,a_2b_2}\right.
\left.+S^{(R)}_{v,a_1b_1}((v\cdot D)S_v^{(R')})_{a_2b_2}
\right] =0.
\end{eqnarray}
Since the Wilson line~\eqref{eq:def-wilson} is the unique solution
of the differential equation~\eqref{eq:diff-wilson} with boundary condition 
$\lim_{s\to \infty}S_v(sv)=1$, eq.~\eqref{eq:combine-wilson} follows.

%%%%%%%%%%%%%%%%%%%%%%%%%%%%%%%%%
\section{Clebsch-Gordan coefficients, colour bases and projectors 
for gluino and squark production}
\label{app:susy-colour}
\subsection{Squark-antisquark (top-antitop) production}
\label{eq:3-3b}
We collect here the Clebsch-Gordan coefficients,
projection operators and basis tensors for quark-antiquark and gluon-gluon initiated production of a $3\otimes \bar 3$ final state. 
The projectors on the singlet and octet final state representations
are independent of the production channel and have been given 
already in~\eqref{eq:projectors-fund}:
\begin{equation}
\begin{aligned}
  P^{(1)}_{\{a\}}&=\frac{1}{N_c}\,\delta_{a_1a_2}\delta_{a_3a_4},\\
 P^{(8)}_{\{a\}}&=2\,T^\alpha_{a_1a_2}T^\alpha_{a_4a_3}.
\end{aligned}
\end{equation}
The indices take the values $a_i\in\{1,2,3\}$.

\subsubsection{Quark-antiquark fusion channel}
For completeness we repeat the results for the quark-antiquark channel 
given already in~\eqref{eq:cgk33} and~\eqref{eq:basis_33}.
The Clebsch-Gordan coefficients for the two
representations in the decomposition $3\otimes\bar 3=1+8$ are
\begin{equation}
\begin{aligned}
  C^{(1)}_{a_1a_2}&=\frac{1}{\sqrt{N_c}}\,\delta_{a_1a_2},\\
   C^{(8)}_{\alpha a_1a_2}&=\sqrt 2 \,T^\alpha_{a_2a_1}
\end{aligned}
\label{eq:cgk33-app}
\end{equation}
with $\alpha\in\{1,\dots,8\}$.
The basis elements for the colour structure of the hard production
process corresponding to the two possible combinations
$P_i=\{(1,1),(8,8)\}$ are given by
\begin{equation}
\begin{aligned}
c^{(1)}_{\{a\}} &= \frac{1}{N_c} \,\delta_{a_1a_2} \delta_{a_3a_4}  \,,\\
c^{(2)}_{\{a\}} &=\frac{2}{\sqrt{D_A}} \,T^\beta_{a_2a_1} T^\beta_{a_3a_4}
\end{aligned}
\end{equation}
with $D_A=N_c^2-1$ and $N_c=3$.

\subsubsection{Gluon fusion channel}
For the production of a $3\otimes \bar 3$ final state from gluon fusion there are three possible combinations of equivalent initial and 
final state representations: 
\begin{equation}
\label{eq:tensor-33}
P_i\in\{(1,1),\; (8_S,8),\;(8_A,8)\}.
\end{equation}
The Clebsch-Gordan coefficients for combining two particles in the adjoint 
into a singlet, a symmetric and an antisymmetric octet are
\begin{equation} \label{eq:cgk83}
\begin{aligned}
C^{(1)}_{a_1a_2} &= \frac{1}{\sqrt{D_A}}\,\delta_{a_1a_2},\\
C^{(8_S)}_{\alpha a_1a_2} &= \frac{1}{2\sqrt{B_F}}\,D^{\alpha}_{ a_2a_1},\\
C^{(8_A)}_{\alpha a_1a_2} &=
\frac{1}{\sqrt{N_c}} \,F^\alpha_{a_2a_1}, 
\end{aligned}
\end{equation}
where all indices run from $1$ to $8$. 
Here we have defined $F^\alpha_{a_1a_2} = i f^{a_1\alpha a_2}$ in terms of 
the SU(3) structure constants, the symmetric invariant tensor
$D^\alpha_{a_1a_2}=d^{\alpha a_1a_2}$, and the coefficient
$B_F=\frac{N_c^2-4}{4 N_c}=\frac{5}{12}$ appearing in the relation
$\tr[D^\alpha D^\beta]=4B_F\delta_{\alpha\beta}$.  From the
definition~\eqref{eq:prod-basis} we obtain the same colour basis
that has been found to diagonalize the one-loop soft anomalous
dimension matrix~\cite{Kidonakis:1997gm} 
\begin{equation} \label{eq:basis_83}
\begin{aligned}
c^{(1)}_{\{a\}} &= \frac{1}{\sqrt{N_c D_A}} \,
\delta_{a_1a_2}\delta_{a_3a_4} ,\\
c^{(2)}_{\{a\}} &= \frac{1}{\sqrt{2 D_A B_F}}\, 
D^\alpha_{a_2a_1}T^\alpha_{a_3a_4}, \\
c^{(3)}_{\{a\}} &=\sqrt{\frac{2}{N_cD_A}} \,F^\alpha_{a_2a_1} T^\alpha_{a_3a_4}.
\end{aligned}
\end{equation}
\subsection{Squark-squark production}
For quark-quark initiated processes $q q \rightarrow \tilde q \tilde
q$ the initial- and final-state systems are either in the $\bar 3$ or
$6$ representation of SU(3) since $3 \otimes 3=\bar{3}+6$.  We
denote the $6$ by a symmetric double index $\alpha=(\alpha_1\alpha_2)$ with
$\alpha_i\in\{1,2,3\}$.  The Clebsch-Gordan coefficients are given by
\begin{equation}
\label{eq:clebsch-3-6}
\begin{aligned}
  C^{(\bar 3)}_{\alpha a_1a_2}&= \frac{1}{\sqrt 2}\epsilon_{\alpha a_1a_2},\\
  C^{(6)}_{\alpha a_1a_2}&=\frac{1}{2}
  (\delta_{\alpha_1a_1}\delta_{\alpha_2a_2}+
\delta_{\alpha_1a_2}\delta_{\alpha_2a_1}),
\end{aligned}
\end{equation}
where the indices $a_i$ and the index $\alpha$ for the case of the $\bar 3$ 
can take the values $1$ to $3$.
For the sextet representation the normalization condition uses a symmetrized
definition of the Kronecker-delta for the double indices:
\begin{equation}
  \delta^{(6)}_{\alpha\beta}=
\frac{1}{2}(\delta_{\alpha_1\beta_1}\delta_{\alpha_2\beta_2}+
\delta_{\alpha_1\beta_2}\delta_{\alpha_2\beta_1}).
\end{equation}
The projectors on the two representations needed for the 
decomposition of the Coulomb Green function are given by
\begin{equation}
\begin{aligned}
P^{(\bar 3)}_{\{a\}} &= \frac{1}{2}
 \left(\delta_{a_1a_3} \delta_{a_2a_4}-\delta_{a_1a_4} 
\delta_{a_2a_3}\right),\\
P^{(6)}_{\{a\}} &= \frac{1}{2}
 \left(\delta_{a_1a_3} \delta_{a_2a_4}+\delta_{a_1a_4} \delta_{a_2a_3}\right).
\end{aligned}
\end{equation}
Since the tensor product $3\otimes 3$ is of two identical representations 
and since the Clebsch-Gordan coefficients are real, the elements of the 
colour basis coincide with the projectors, up to normalization:
\begin{equation}
\begin{aligned}
\label{eq:basis_33-s}
c^{(1)}_{\{a\}} &= \frac{1}{\sqrt{2N_c(N_c-1)}}
 \left(\delta_{a_1a_3} \delta_{a_2a_4}-
\delta_{a_1a_4} \delta_{a_2a_3}\right),\\
c^{(2)}_{\{a\}} &= \frac{1}{\sqrt{2N_c(N_c+1)}}
 \left(\delta_{a_1a_3} \delta_{a_2a_4}+\delta_{a_1a_4} \delta_{a_2a_3}\right).
\end{aligned}
\end{equation}

%%%%%%%%%%%%%%%%%%%%%%%%%%%%%%%%%%%%%%%%%%%%%%%%%%%%%%%%
\subsection{Gluino-squark production}
For gluino-squark production $qg\to \tilde q\tilde g$
the relevant representations appear in
the decomposition $3\otimes 8=3+\bar 6 +15$. 
The Clebsch-Gordan coefficients are given by (for $N_c=3$)
\begin{equation}
\begin{aligned}
  C^{(3)}_{\alpha a_1a_2}&= \frac{1}{\sqrt{C_F}}T^{a_2}_{\alpha a_1} ,\\
  C^{(\bar 6)}_{\alpha a_1a_2}&=
  \frac{1}{2}(\epsilon_{\alpha_1b a_1}T^{a_2}_{b\alpha_2}
  +\epsilon_{\alpha_2b a_1}T^{a_2}_{b\alpha_1}),&\\
 C^{(15)}_{\alpha a_1a_2}&=
  \frac{1}{\sqrt 2}\left(\delta_{\alpha_1 a_1}T^{a_2}_{\alpha_2\alpha_3}
  +\delta_{\alpha_2 a_1}T^{a_2}_{\alpha_1\alpha_3} 
  -\frac{1}{4}\delta_{\alpha_1 \alpha_3}T^{a_2}_{\alpha_2a_1}
  -\frac{1}{4}\delta_{\alpha_2 \alpha_3}T^{a_2}_{\alpha_1a_1}\right), 
\end{aligned}
\end{equation}
where $a_1\in\{1,2,3\}$ and  $a_2\in\{1,\dots,8\}$. 
For the $\bar 6$ we use the same double-index convention as
in~\eqref{eq:clebsch-3-6}. For the $15$ we have introduced a triple
index $\alpha=(\alpha_1\alpha_2\alpha_3)$ where the first two
indices transform in the $3$ and the last index transforms in the
$\bar 3$ representation. The Clebsch-Gordan coefficient is symmetric
under the exchange $\alpha_1\leftrightarrow\alpha_2$ and vanishes upon
contracting $\alpha_{1,2}$ with $\alpha_3$.  In the normalization of
the coefficient $ C^{(15)}$ we use a Kronecker delta that has the same
symmetries as the coefficient in both index triples:
\begin{equation}
  \delta^{(15)}_{\alpha\beta}=
\frac{1}{2}
\delta_{\alpha_1\beta_1}\delta_{\alpha_2\beta_2}\delta_{\alpha_3\beta_3}
-\frac{1}{8}\delta_{\alpha_1\alpha_3}\delta_{\alpha_2\beta_2}\delta_{\beta_1\beta_3}
-\frac{1}{8}\delta_{\alpha_1\alpha_3}\delta_{\alpha_2\beta_1}\delta_{\beta_2\beta_3}
+(\alpha_1\leftrightarrow \alpha_2 ).
\end{equation}

The projectors can be written as (for $N_c=3$)
\begin{equation}
\label{eq:P38-38}
\begin{aligned}
P^{(3)}_{\{a\}}&=\frac{1}{C_F}\,(T^{a_2}T^{a_4})_{a_1a_3},\\
P^{(\bar 6)}_{\{a\}}&=
\frac{1}{2}\delta_{a_1a_3}\delta_{a_2a_4}
-\frac{1}{2}(T^{a_2}T^{a_4})_{a_1a_3}
-(T^{a_4}T^{a_2})_{a_1a_3},\\
P^{(15)}_{\{a\}}
&=\frac{1}{2}\delta_{a_1a_3}\delta_{a_2a_4}
-\frac{1}{4}(T^{a_2}T^{a_4})_{a_1a_3}
+(T^{a_4}T^{a_2})_{a_1a_3}.
\end{aligned}
\end{equation}
The colour basis that diagonalizes the one-loop soft function is
related to the projectors by complex conjugation and a normalization
factor $1/\sqrt{\text{dim}(r_\alpha)}$:
\begin{eqnarray}
c^{(1)}_{\{a\}}&=&\frac{1}{\sqrt 3C_F}(T^{a_4}T^{a_2})_{a_3a_1},
\nonumber\\
c^{(2)}_{\{a\}}
&=&\frac{1}{2\sqrt 6}\left(\delta_{a_1a_3}\delta_{a_2a_4}
-\,(T^{a_4}T^{a_2})_{a_3a_1}-2 (T^{a_2}T^{a_4})_{a_3a_1}\right),\\
c^{(3)}_{\{a\}}
&=&\frac{1}{\sqrt{15}}\left(\frac{1}{2}\delta_{a_1a_3}\delta_{a_2a_4}
-\frac{1}{4}(T^{a_4}T^{a_2})_{a_3a_1}
+(T^{a_2}T^{a_4})_{a_3a_1}\right).
\nonumber
\end{eqnarray}

%%%%%%%%%%%%%%%%%%%%%%%
\subsection{Gluino pair production}
 The projectors on the several final-state representations for 
gluino pairs appearing in the decomposition of $8\otimes 8$ are the same  
 for the quark-antiquark and gluon-induced processes and can be
 obtained from~\cite{MacFarlane:1968vc,Dokshitzer:2005ig}:
\begin{equation}
\begin{aligned}
P^{(1)}_{\{a\}} &= 
\frac{1}{8} \delta_{a_1 a_2} \delta_{a_3 a_4},\\
P^{(8_S)}_{\{a\}} &= 
\frac{3}{5} D^\alpha_{a_1 a_2}D^\alpha_{a_4a_3 },\\
P^{(8_A)}_{\{a\}} &= 
\frac{1}{3} F^\alpha_{a_1 a_2} F^\alpha_{a_4 a_3}, \\
P^{(10)}_{\{a\}}&=\frac{1}{4}\left(\delta_{a_1 a_3}\delta_{a_2 a_4}-
\delta_{a_1 a_4} \delta_{a_2 a_4}
-\frac{2}{3} F^\alpha_{a_1 a_2} F^\alpha_{a_4 a_3} +
D^\alpha_{a_3 a_1} F^\alpha_{a_4 a_2}+F^\alpha_{a_3 a_1} D^\alpha_{a_4 a_2}\right),
\\
P^{(\overline{10})}_{\{a\}}&
=\frac{1}{4}\left(\delta_{a_1 a_3}\delta_{a_2 a_4}-
\delta_{a_1 a_4} \delta_{a_2 a_4}
-\frac{2}{3} F^\alpha_{a_1 a_2} F^\alpha_{a_4 a_3} -D^\alpha_{a_3 a_1} F^\alpha_{a_4 a_2}
-F^\alpha_{a_3 a_1} D^\alpha_{a_4 a_2}\right),
\\
P^{(27)}_{\{a\}} &= \frac{1}{2} \left(\delta_{a_1 a_3} \delta_{a_2 a_4}
+\delta_{a_1 a_4} \delta_{a_2 a_3}
-\frac{1}{4} \delta_{a_1 a_2} \delta_{a_3 a_4}
-\frac{6}{5} D^\alpha_{a_2 a_1} D^\alpha_{a_3 a_4} \right).
\end{aligned}
\label{eq:project-ad}
\end{equation}
\subsubsection{Quark-antiquark fusion channel}
For quark-antiquark induced processes, the pairs of equivalent 
combinations of initial and final state representations are
\begin{equation}
\label{eq:tensor-33-88}
P_i\in\{(1,1),\; (8,8_S),\;(8,8_A)\}.
\end{equation} The Clebsch-Gordan coefficients that
combine the initial state quarks into a singlet and an octet have been
given in~\eqref{eq:cgk33-app} while the coefficients for the final state
appeared in~\eqref{eq:cgk83}.  The resulting colour basis is
related to~\eqref{eq:basis_83} by exchanging initial and final states:
\begin{equation}
\label{eq:basis_38} 
\begin{aligned} 
c^{(1)}_{\{a\}} &= \frac{1}{\sqrt{N_c D_A}}\,\delta_{a_1a_2}\delta_{a_3a_4},
\\
c^{(2)}_{\{a\}} &= \frac{1}{\sqrt{2 D_A B_F}} \,T^\alpha_{a_2a_1} D^\alpha_{a_3a_4},
\\
c^{(3)}_{\{a\}} &= 
\sqrt{\frac{2}{N_cD_A}} \,T^\alpha_{a_2a_1}  F^\alpha_{a_3a_4}.
\end{aligned}
\end{equation}

\subsubsection{Gluon fusion channel}
For the production of two gluinos from gluon fusion, the allowed pairs 
of initial and final state representations are given by~\eqref{eq:tensor-88}:
\begin{equation}
P_i\in 
\{(1,1),\; (8_S,8_S),\;(8_A,8_S),\;(8_A,8_A),\;(8_S,8_A),\;\;(10,10),\;
(\overline{10},\overline{10}),\; (27,27) \}.
\end{equation}
The Clebsch-Gordan coefficients for the singlet and octet
representations are the same as in~\eqref{eq:cgk83}. Since according 
to~\eqref{eq:prod-basis} the colour
basis elements for the production of a $10$, $\overline{10}$ and $27$
 are, up to normalization, the complex conjugate of the projectors given 
in~\eqref{eq:project-ad}, we do not need the lengthy Clebsch-Gordan 
coefficients for these representations. The
basis consists of an operator corresponding to the production of a
singlet:
\begin{equation}
c^{(1)}_{\{a\}} = 
\frac{1}{D_A}\,
 \delta_{a_1a_2} \delta_{a_3a_4} ,
\end{equation}
four operators corresponding to the different combinations of $8_S$ and $8_A$,
\begin{equation}
\begin{aligned}
\label{eq:basis-88-octet}
c^{(2)}_{\{a\}} &= \frac{1}{4B_F\sqrt{D_A}}\,
D^\alpha_{a_2 a_1} D^\alpha_{a_3 a_4}, \\
c^{(3)}_{\{a\}} &= 
\frac{1}{2\sqrt{B_F N_cD_A}}\,
F^\alpha_{a_2 a_1} D^{\alpha}_{a_3 a_4}, \\
c^{(4)}_{\{a\}} &= \frac{1}{N_c\sqrt{D_A}}\,
F^\alpha_{a_2 a_1} F^\alpha_{a_3 a_4} ,\\
c^{(5)}_{\{a\}} &= \frac{1}{2\sqrt{B_F N_c D_A}}\,
D^\alpha_{a_2a_1} F^\alpha_{a_3a_4},
\end{aligned}
\end{equation}
two operators corresponding to the $10$ and $\overline{10}$
\begin{equation}
c^{(6/7)}_{\{a\}} = \frac{ 1}{4 \sqrt{10}} 
\left[\delta_{a_1 a_3}\delta_{a_2 a_4}-\delta_{a_1 a_4} \delta_{a_2 a_4}
-\frac{2}{3} F^\alpha_{a_2 a_1} F^\alpha_{a_3 a_4}  
\pm
\left(D^\alpha_{a_3 a_1} F^\alpha_{a_4 a_2}+F^\alpha_{a_3 a_1} D^\alpha_{a_4 a_2}\right)\right],
\end{equation}
and one operator for the $27$:
\begin{equation}
c^{(8)}_{\{a\}} = \frac{1}{6 \sqrt{3}} \left(\delta_{a_1a_3} \delta_{a_2 a_4}
+\delta_{a_1 a_4} \delta_{a_2 a_3}
-\frac{1}{4} \delta_{a_1a_2} \delta_{a_3a_4}
-\frac{6}{5} D^\alpha_{a_2 a_1} D^\alpha_{a_3 a_4} \right).
\end{equation}

%%%%%%%%%%%%%%%%%%%%%%%%%%%%%%%%%%%%%%%%%%%%%%%%%%%%%%%%%%%%%%%%%%%%%%%%%
\section{Fourier transform of the soft function}
\label{app:fourier}

The Fourier transform of the soft function~\eqref{eq:soft-ft} enters
the factorization formula~\eqref{eq:fact} while in section~\ref{sec:soft-one} we calculated the one-loop soft function in position space.
The momentum space result can be obtained by inserting the Fourier transforms of the basis integrals 
\begin{eqnarray}
 \mathcal{I}^{(\text{ii})}(\omega,\mu) &=& 
-\frac{\Gamma(-\epsilon)}{8\pi^{2}\Gamma(-2\epsilon)}\frac{1}{\epsilon}
e^{\gamma_E\epsilon} \;\frac{1}{\omega}\left(\frac{\omega}{\mu}\right)^{-2\epsilon}
\theta\left(\omega\right),
\\[0.2cm]
\mathcal{I}^{(\text{ff})}(\omega,\mu)&=&
\frac{\Gamma(-\epsilon)}{8\pi^{2}\Gamma(-2\epsilon)} e^{\gamma_E\epsilon}
\frac{1}{(1-2\epsilon)} \frac{1}{\omega}
\left(\frac{\omega}{\mu}\right)^{-2\epsilon}
\theta\left(\omega\right)\nonumber 
\end{eqnarray}
and $\mathcal{I}^{(\text{if})}(\omega,\mu)=-1/2\;
\mathcal{I}^{(\text{ii})}(\omega,\mu)$ into the
result~\eqref{eq:soft-result-1}.   For unstable heavy particles, the
$\omega$ integral in~\eqref{eq:fact} extends to infinity and the
 integrals have to be expanded in $\epsilon$  in the
sense of modified plus-distributions~\cite{Actis:2008rb}. 
 The Fourier
transform has been calculated using 
\begin{equation}
  \int_{-\infty}^{\infty}\frac{dz_0}{4\pi} e^{i\omega z_0/2}
 \left(\frac{iz_0\mu}{2}\right)^\alpha =
\frac{1}{\Gamma(-\alpha)} 
\frac{1}{\omega}\left(\frac{\omega}{\mu}\right)^{-\alpha} \theta(\omega),
\end{equation}
where we recall the prescription $z_0\to z_0-i\delta$.

To compare the one-loop soft function~\eqref{eq:soft-one} to
the result obtained in~\cite{Idilbi:2009cc} for the special case of
the production of a colour octet scalar from gluon fusion, 
we compute the Fourier transform as a function of $\omega=M
(1-z)$ and expand the basis integrals into plus-distributions, as
appropriate if the decay width is neglected as in~\cite{Idilbi:2009cc}:
\begin{eqnarray}
 \mathcal{I}^{(\text{ii})}(M(1-z),\mu)
&=&\frac{1}{8 \pi^2 M}\left[\delta(1-z)\left(\frac{1}{\epsilon^2}
  +\frac{2}{\epsilon}\ln\left(\frac{\mu}{M}\right)+
  2\ln^2\left(\frac{\mu}{M}\right)-\frac{\pi^2}{4} \right)\right.
\nonumber\\
&&\left.  -\,\left[\frac{1}{1-z}\right]_+\left(\frac{2}{\epsilon}
  +4\ln\left(\frac{\mu}{M}\right)
\right)
+4\left[\frac{\ln(1-z)}{1-z}\right]_+
\right]\theta(1-z),
\\[0.2cm]
\mathcal{I}^{(\text{ff})}(M(1-z),\mu)
&=&-\frac{1}{8 \pi^2 M}\left[\delta(1-z)\left(\frac{1}{\epsilon}
  +2\ln\left(\frac{\mu}{M}\right)+2\right)
 - 2 \left[\frac{1}{1-z}\right]_+\right]\theta(1-z).
\nonumber
\end{eqnarray}
Here we have  used the identity 
\begin{equation}
  (1-z)^{-1-2\epsilon}=-\frac{1}{2\epsilon}\delta(1-z)
  +\left[\frac{1}{1-z}\right]_+-2\epsilon\left[\frac{\ln(1-z)}{1-z}\right]_+ 
\end{equation}
for distributions on the interval $[0,1]$.

The soft function in momentum space obtained from~\eqref{eq:soft-one} 
is therefore given by
\begin{eqnarray}
\label{eq:soft-one-ft}
 W^{(1)R_\alpha}_{i}(M(1-z))&=&
 \frac{2}{M}\left\{\left(C_r+C_{r'}\right)
\left[\delta(1-z)\left(\frac{1}{\epsilon^2}
  +\frac{2}{\epsilon}\ln\left(\frac{\mu}{M}\right)+
  2\ln^2\left(\frac{\mu}{M}\right)-\frac{\pi^2}{4} \right)\right.\right . 
\nonumber\\
&& \left.\left.  -\,\left[\frac{1}{1-z}\right]_+\left(\frac{2}{\epsilon}
  +4\ln\left(\frac{\mu}{M}\right)\right)
+4\left[\frac{\ln(1-z)}{1-z}\right]_+\right]\right.
\nonumber\\
&&\left.+\,C_{R_\alpha}\left[\delta(1-z)\left(\frac{1}{\epsilon}
  +2\ln\left(\frac{\mu}{M}\right)+2\right)
 - 2\left[\frac{1}{1-z}\right]_+\right]\right\}.
\end{eqnarray}
This reproduces eq.~(40) in~\cite{Idilbi:2009cc} by
setting $C_r=C_{r'}=C_{R_\alpha}=C_A=N_c$ and multiplying with a
prefactor $\alpha_sM/(4\pi)$ to account for our definition of
$W^{(1)R_\alpha}_i$ as coefficient of $\alpha_s/(4\pi)$ and the different
normalization of the leading-order soft function:
$W^{(0)R_{\alpha}}_i(M(1-z))=\delta(M(1-z))=1/M \delta(1-z)$, whereas
$\overline{S}^{(0)}_{S/P}(M(1-z))=\delta(1-z)$
in~\cite{Idilbi:2009cc}.

%%%%%%%%%%%%%%%%%%%%%%%%%%%%%%%%%%%%%%%%

\section{Anomalous dimensions}
\label{app:gamma}

In~\ref{app:gamma-hard} we derive
the anomalous dimension~\eqref{eq:gamma-sc} of the hard function. 
The explicit one- and two-loop results for the anomalous dimensions are
collected in~\ref{app:gamma-results}.

\subsection{Anomalous-dimension matrix of the hard function}
\label{app:gamma-hard}

 From~\cite{Becher:2009kw} we find that the UV regularized, minimally
 subtracted short-distance coefficients for a general $2\to n$
 scattering process including massless and massive partons obey
 an evolution equation
\begin{equation}
\frac{d}{d\ln\mu}\ket{\mathcal{C}(\{k\},\{m\},\mu)}= 
{\bf \Gamma}(\{k\},\{m\},\mu)\ket{\mathcal{C}(\{k\},\{m\},\mu)},
\end{equation}
where $\{k\}=\{k_1,\dots,k_n\}$ and $\{m\}=\{m_1,\dots,m_n\}$ denote momenta
and masses of the scattered particles and we use the colour-state
formalism~\cite{Catani:1996vz}.  For general scattering processes the colour
structure of the matrix ${\bf \Gamma}$ can be quite complicated with two- and
three-parton colour correlations contributing at two-loop level.  

The two-parton correlations take the form (c.f.~eq.(10) 
of \cite{Becher:2009kw}) 
\begin{eqnarray}
\label{eq:gamma}
  {\bf \Gamma}(\{k\},\{m\},\mu)|_{\text{2-parton}} &=&\sum_{(i,j)}
  \frac{{\bf T}_i\cdot {\bf T}_j}{2}\; \gamma_{\text{cusp}}\;
  \ln\left(\frac{\mu^2}{-s_{ij}}\right)+\sum_i \gamma^{r_i}
+\sum_I\gamma_{H,s}^{R_I}
\nonumber \\
&& \hspace*{-3cm} 
- \,\sum_{(I,J)} \frac{{\bf T}_I\cdot {\bf T}_J}{2}\,
\gamma_{\text{cusp}}(\beta_{IJ})
+ \sum_{I,j}{\bf T}_I\cdot {\bf T}_j\,
\gamma_{\text{cusp}} \,\ln\left(\frac{m_J\mu}{-s_{Ij}}\right).
\end{eqnarray}
Here indices $i$~($I$) denote massless (massive) partons and the
notation $(i,j)$ indicates unordered tuples of distinct parton
indices.  We also defined $s_{ij}=2\sigma_{ij} p_i\cdot p_j+i 0$ with
$\sigma_{ij}=+1$ if partons $i$ and $j$ are both incoming or outgoing
and $\sigma_{ij}=-1$ otherwise.

The anomalous dimensions related to light partons are collected in
appendix~\ref{app:gamma-results}.  
The  heavy-particle soft anomalous dimension $\gamma_{H,s}^{R_I}$ for 
arbitrary SU(3) representations is given in~\eqref{eq:gamma-H}.
The cusp anomalous dimension
$\gamma_{\text{cusp}}(\beta_{IJ})$ in the terms involving two massive partons
is a function of the cusp angle $\beta_{IJ}=\text{arccosh}(-s_{IJ}/2m_Im_J)$
and can be obtained from the heavy-heavy quark formfactor in
HQET~\eqref{eq:heavy}. For large cusp angles the cusp anomalous dimension
satisfies~\eqref{eq:cusp-infty} while for small cusp angle as in pair
production at threshold~\cite{Becher:2009kw}
\begin{equation}
\label{eq:cusp-threshold}
  \gamma_{\text{cusp}}(\beta)\xrightarrow{\beta\to 0} 
  -2\gamma^{Q}/C_{F}\equiv \gamma_{H,s}.
\end{equation}

For processes with two heavy final state particles three-parton
correlations involve the colour structures 
\begin{equation}
\label{eq:three-parton}
f^{abc}\,{\bf T}^a_i{\bf
  T}^b_J{\bf T}^c_K,
\end{equation}
while a two-loop analysis~\cite{Mitov:2009sv} and soft-collinear
factorization~\cite{Becher:2009kw} show that three-parton correlations
with two light partons and one heavy parton proportional to
$f^{abc}\,{\bf T}^a_i{\bf T}^b_j{\bf T}^c_K$ are absent. 

In the following we evaluate~\eqref{eq:gamma} for a $2\to 1$ process
with a massive final state particle in an arbitrary SU(3) representation,
corresponding to the reduction of the leading soft function obtained in
section~\ref{sec:colour}. As discussed in section~\ref{sec:nnll} this
corresponds to dropping contributions to the anomalous dimension
related to higher-order potentials and higher-dimensional soft
functions.  For the three-parton process with a single heavy particle
the structure (\ref{eq:three-parton}) cannot appear.  Furthermore, for
a two-to-one process the structure $f^{abc}\,{\bf T}^a_i{\bf
  T}^b_j{\bf T}^c_K$ vanishes by colour conservation alone.  We
therefore conclude that in order to obtain the two-loop anomalous
dimension of the hard function in the factorization formula~\eqref{eq:fact}
corresponding to the (leading) soft function $W_{i i'}^{R_\alpha}$ 
in the same equation, it is sufficient to consider the two-parton
correlations~\eqref{eq:gamma}, since all higher multi-particle
correlations do not contribute.

The result~\eqref{eq:gamma-sc} for the anomalous dimension of the hard
coefficient in heavy particle pair production at threshold 
can be obtained from the
general expression~\eqref{eq:gamma} using the reduction to a two-to-one
process with a single particle in the representation $R_\alpha$ with mass
$2M=(m_H+m_{H'})$ and momentum $P=p_1+p_2$, derived for the soft function in
section~\ref{sec:colour}. (This reduction can be performed analogously for the
soft function $\braket{0|S_nS_{\bar n}S_v^\dagger S_v^\dagger|0}$ appearing in
the factorization of the amplitude to which eq.~\eqref{eq:gamma} applies.) 
At threshold $k_1\cdot k_2=P\cdot
k_i=2M^2$ and the two-loop anomalous-dimension matrix becomes
\begin{eqnarray}
&& {\bf \Gamma}(\{k,P\},M,\mu) = 
  {\bf T}_1\cdot {\bf T}_2\; \gamma_{\text{cusp}}\;
  \ln\left(-\frac{\mu^2}{2k_1\cdot k_2+i 0}\right)
+\gamma^r+\gamma^{r'}+\gamma_{H,s}^{R_\alpha} 
\nonumber\\
&& \hspace*{2cm} 
+ \,\sum_{i=1,2}{\bf T}_i\cdot {\bf T}_3\,\gamma_{\text{cusp}}
\ln\left(\frac{2M\mu}{2(P\cdot k_i)}\right) \\
&& \hspace*{2cm} = \,\gamma_{\text{cusp}}\left[
  {\bf T}_1\cdot {\bf T}_2\,
 \ln\left(-\frac{\mu^2}{4M^2+i 0}\right)
 -{\bf T}_3^2\; \ln\left(\frac{\mu}{2M}\right)\right]
+\gamma^r+\gamma^{r'}+\gamma^{R_\alpha}_{H,s},\quad
\nonumber
\end{eqnarray}
where we have used colour conservation 
${\bf T}_1+{\bf T}_2=-{\bf T}_3$. 
Further using the identity
\begin{equation}
\label{eq:t12}
2 \,{\bf T}_1\cdot {\bf T}_2={\bf T}_3^2-{\bf T}_1^2-{\bf T}_2^2 
\end{equation}
leads to the result for the anomalous dimension  quoted
in~\eqref{eq:gamma-sc}:
\begin{equation}
{\bf \Gamma}(\{k,P\},M,\mu)
=\frac{1}{2}\,\gamma_{\text{cusp}}\left[
(C_r+C_{r'})\left(\ln\left(\frac{4M^2}{\mu^2}\right)-i \pi\right)
+i\pi C_{R_\alpha}\right]
+\gamma^r+\gamma^{r'}+\gamma_{H,s}^{R_\alpha}.
\end{equation}

It is interesting to see how the same result is obtained by applying the
general expression~\eqref{eq:gamma} directly to the four-particle process at
threshold, omitting three parton correlations as discussed above.
 Using the fact that the heavy-particle cusp anomalous dimension
near threshold simplifies according to~\eqref{eq:cusp-threshold} 
and $p_J\cdot k_i=m_J\sqrt{\hat s}/2\approx m_JM$ one finds
\begin{eqnarray}
  {\bf \Gamma}(\{k,p\},\{m\},\mu)|_{\text{2-parton}}&=&
  {\bf T}_1\cdot {\bf T}_2\, \gamma_{\text{cusp}}\,
  \ln\left(-\frac{\mu^2}{2 k_1\cdot k_2+i0}\right)+\gamma^r+\gamma^{r'}
+\gamma_{H,s}^{R}+\gamma_{H,s}^{R'}
\nonumber \\
&& +\,2 \,{\bf T}_3\cdot {\bf T}_4\,\gamma_{H,s}
+ \sum_{i,J=1,2}{\bf T}_i\cdot {\bf T}_{J+2}\,\gamma_{\text{cusp}}\,
\ln\left(\frac{m_J\mu}{2 p_J\cdot k_i}\right)
\nonumber \\
&=& {\bf T}_1\cdot {\bf T}_2\, \gamma_{\text{cusp}}\,
  \ln\left(-\frac{\mu^2}{4M^2+i 0}\right)
+\gamma^r+\gamma^{r'}+2 \,{\bf T}_3\cdot {\bf T}_4\,\gamma_{H,s}\quad
\nonumber \\
&& +\,({\bf T}_1+ {\bf T}_2)\cdot({\bf T}_3+ {\bf T}_4)\,\gamma_{\text{cusp}}
\ln\left(\frac{\mu}{2M}\right)+\gamma_{H,s}^{R}+\gamma_{H,s}^{R'}.
\end{eqnarray}
Using colour conservation and the analog of~\eqref{eq:t12} for a
four-particle process, this expression simplifies to
\begin{eqnarray}
 {\bf \Gamma}(\{k,p\},\{m\},\mu)|_{\text{2-parton}}
&=& \frac{1}{2}\,\gamma_{\text{cusp}}\,\left[({\bf T}_1^2+ {\bf T}_2^2)\,
  \left(\ln\left(\frac{4M^2}{\mu^2}\right)-i\pi\right)
  +i\pi({\bf T}_3+{\bf T}_4)^2\right]\nonumber\\[0.2cm]
&& +\,\gamma^r+\gamma^{r'}
  + \left[({\bf T}_3+ {\bf T}_4)^2-{\bf T}_3^2- {\bf T}_4^2\right]\gamma_{H,s}
+\gamma_{H,s}^{R}+\gamma_{H,s}^{R'}.\quad
\end{eqnarray}
Since for a final state pair in a representation
$R_\alpha$, $({\bf T}_3+{\bf T}_4)^2=C_{R_\alpha}$, we obtain the same
result as in the three-point calculation, provided the heavy-particle
soft anomalous dimension satisfies Casimir scaling,
$\gamma^{R_\alpha}_{H,s}=C_{R_\alpha}\gamma^H_s$.  This gives a second
argument for Casimir scaling of the soft anomalous dimension, in
addition to the derivation from the HQET formfactor given in the main
text.

\subsection{Explicit results for the anomalous dimensions}
\label{app:gamma-results}
In this appendix we collect explicit results for the one- and two-loop
anomalous dimensions that are already available in the literature.
We define the expansion of the various
anomalous dimensions in the strong coupling constant by
\begin{equation}\label{eq:gamma-alpha}
\gamma=\sum_n \gamma^{(n)}
\left(\frac{\alpha}{4\pi}\right)^{n+1}.
\end{equation}
The explicit one- and two-loop results for the cusp anomalous dimension are
given by
\begin{equation}
\label{eq:cusp}
\begin{aligned}
\gamma^{(0)}_{\text{cusp}}&= 4 ,\\
\gamma^{(1)}_{\text{cusp}}&= 4\left[
\left(\frac{67}{9}-\frac{\pi^2}{3}\right)C_A
-\frac{20}{9} T_F n_f\right]%    \\\;
\end{aligned}
\end{equation}
with $T_F\delta_{ab}=\tr[T^aT^b]$ and $T_F=1/2$.

The one- and two-loop
anomalous-dimension coefficients $\gamma^r$ of massless quarks,
$\gamma^q=\gamma^3$, and gluons, $\gamma^g=\gamma^8$, are given
by~\cite{Becher:2009qa}
\begin{align}
 \gamma^{(0)q}=&-3C_F ,\\
 \gamma^{(1)q}=&\;C_F^2\left(-\frac{3}{2}+2\pi^2-24\zeta_3\right)
   +C_AC_F\left(-\frac{961}{54}-\frac{11\pi^2}{6}+26\zeta_3\right)
  \nonumber\\ 
   &+C_FT_F n_f\left(\frac{130}{27}+\frac{2\pi^2}{3}\right),\\
 \gamma^{(0)g}=&-\beta_0 =-\frac{11}{3} C_A+\frac{4}{3} T_F n_f ,\\
 \gamma^{(1)g}=&\;C_A^2\left(-\frac{692}{27}+\frac{11\pi^2}{18}+
     2\zeta_3\right)
    +C_AT_F n_f\left(\frac{256}{27}-\frac{2\pi^2}{9} \right)
    +4C_FT_F n_f.
\end{align}

The anomalous dimensions $\gamma^{\phi,r}$ appearing in the anomalous
dimension of the soft function~\eqref{eq:relation-soft-gamma} are defined by
the evolution equation of the parton distribution function for a parton $p$ in
the representation $r$ in the $x\to 1$ limit,
\begin{equation}
\label{eq:def-gamma-phi}
\frac{d}{d\ln\mu}f_{p/N}(x,\mu)
=2 \gamma^{\phi,r}(\alpha_s) f_{p/N}(x,\mu)+
2 \,\Gamma^r_{\text{\text{cusp}}}(\alpha_s)
\int_x^1\frac{dz}{z}\frac{f_{p/N}(x/z,\mu)}{[1-z]_+}
+... \, .
\end{equation}
Results for quarks, $\gamma^{\phi,3}\equiv\gamma^\phi$, and gluons,
$\gamma^{\phi,8}\equiv\gamma^B$, are available up to the three-loop
order~\cite{Moch:2004pa}.  The explicit values in the notation used
here are given by~\cite{Becher:2007ty,Ahrens:2008nc}:
\begin{eqnarray}
  \gamma^{(0)\phi}&=&3C_F ,\\
 \gamma^{(1)\phi}&=&C_F^2\left(\frac{3}{2}-2\pi^2+24\zeta_3\right)
   +C_AC_F\left(\frac{17}{6}+\frac{22\pi^2}{9}-12\zeta_3\right)\nonumber\\ 
   && -\,C_F T_F n_f\left(\frac{2}{3}+\frac{8\pi^2}{9}\right),\\
  \gamma^{(0)B}&=&\beta_0 =\frac{11}{3} C_A- \frac{4}{3} T_F n_f ,\\
 \gamma^{(1)B}&=&4C_A^2\left(\frac{8}{3}+3\zeta_3\right)
    -\frac{16}{3}C_AT_F n_f- 4C_FT_F n_f.
\end{eqnarray}

%%%%%%%%%%%%%%%%%%%%%%%%%%%%%%%%%%%%%%%%%%%%%%%%%%%%%%%%%%%%%%%%%%%%%%%%%%

%%%%%%%%%%%%%%%%%%%%%%%%%%%%%%%%%%%%%%%%%%%%%%%%%%%%%%%%%%%%%%%%%%%%%%%%%%%%%


\begin{thebibliography}{10}

\bibitem{Sterman:1986aj}
G.~Sterman,
\newblock Nucl. Phys. {\bf B281}, 310 (1987).
%%CITATION = NUPHA,B281,310;%%

\bibitem{Catani:1989ne}
S.~Catani and L.~Trentadue,
\newblock Nucl. Phys. {\bf B327}, 323 (1989).
%%CITATION = NUPHA,B327,323;%%

\bibitem{Kidonakis:1997gm}
N.~Kidonakis and G.~Sterman,
\newblock Nucl. Phys. {\bf B505}, 321 (1997), [hep-ph/9705234].
%%CITATION = HEP-PH/9705234;%%

\bibitem{Kidonakis:1998nf}
N.~Kidonakis, G.~Oderda and G.~Sterman,
\newblock Nucl. Phys. {\bf B531}, 365 (1998), [hep-ph/9803241].
%%CITATION = HEP-PH/9803241;%%

\bibitem{Bonciani:2003nt}
R.~Bonciani, S.~Catani, M.~L. Mangano and P.~Nason,
\newblock Phys. Lett. {\bf B575}, 268 (2003), [hep-ph/0307035].
%%CITATION = HEP-PH/0307035;%%

\bibitem{Becher:2006nr}
T.~Becher and M.~Neubert,
\newblock Phys. Rev. Lett. {\bf 97}, 082001 (2006), [hep-ph/0605050].
%%CITATION = HEP-PH/0605050;%%

\bibitem{Becher:2006mr}
T.~Becher, M.~Neubert and B.~D. Pecjak,
\newblock JHEP {\bf 01}, 076 (2007), [hep-ph/0607228].
%%CITATION = HEP-PH/0607228;%%

\bibitem{Becher:2007ty}
T.~Becher, M.~Neubert, and G.~Xu,
\newblock JHEP {\bf 07}, 030 (2008), arXiv:0710.0680 [hep-ph].
%%CITATION = 0710.0680;%%

\bibitem{Catani:1996yz}
S.~Catani, M.~L. Mangano, P.~Nason and L.~Trentadue,
\newblock Nucl. Phys. {\bf B478}, 273 (1996), [hep-ph/9604351].
%%CITATION = HEP-PH/9604351;%%

\bibitem{Bonciani:1998vc}
R.~Bonciani, S.~Catani, M.~L. Mangano and P.~Nason,
\newblock Nucl. Phys. {\bf B529}, 424 (1998), [hep-ph/9801375].
%%CITATION = HEP-PH/9801375;%%

\bibitem{Moch:2008qy}
S.~Moch and P.~Uwer,
\newblock Phys. Rev. {\bf D78}, 034003 (2008), arXiv:0804.1476 [hep-ph].
%%CITATION = 0804.1476;%%

\bibitem{Cacciari:2008zb}
M.~Cacciari, S.~Frixione, M.~L. Mangano, P.~Nason, and G.~Ridolfi,
\newblock JHEP {\bf 09}, 127 (2008), arXiv:0804.2800 [hep-ph].
%%CITATION = 0804.2800;%%

\bibitem{Kidonakis:2008mu}
N.~Kidonakis and R.~Vogt,
\newblock Phys. Rev. {\bf D78}, 074005 (2008), arXiv:0805.3844 [hep-ph].
%%CITATION = 0805.3844;%%

\bibitem{Czakon:2008cx}
M.~Czakon and A.~Mitov,  
Phys.\ Lett.\  B {\bf 680} (2009) 154,
arXiv:0812.0353 [hep-ph].
  %%CITATION = ARXIV:0812.0353;%%

\bibitem{Langenfeld:2009wd}
  U.~Langenfeld, S.~Moch and P.~Uwer,
  Phys.\ Rev.\  D {\bf 80} (2009) 054009,
  arXiv:0906.5273 [hep-ph].
  %%CITATION = PHRVA,D80,054009;%%

\bibitem{Hagiwara:2008df}
K.~Hagiwara, Y.~Sumino, and H.~Yokoya,
\newblock Phys. Lett. {\bf B666}, 71 (2008), arXiv:0804.1014 [hep-ph].
%%CITATION = ARXIV:0804.1014;%%

\bibitem{Kiyo:2008bv}
Y.~Kiyo, J.~H. K{\"u}hn, S.~Moch, M.~Steinhauser and P.~Uwer,
\newblock Eur. Phys. J. {\bf C60}, 375 (2009),  arXiv:0812.0919 [hep-ph].
%%CITATION = 0812.0919;%%

\bibitem{Kulesza:2008jb}
A.~Kulesza and L.~Motyka,
\newblock Phys. Rev. Lett. {\bf 102}, 111802 (2009), arXiv:0807.2405 [hep-ph].
%%CITATION = ARXIV:0807.2405;%%

\bibitem{Kulesza:2009kq}
A.~Kulesza and L.~Motyka,
\newblock (2009), arXiv:0905.4749 [hep-ph].
%%CITATION = ARXIV:0905.4749;%%

\bibitem{Langenfeld:2009eg}
U.~Langenfeld and S.-O. Moch,
\newblock (2009), arXiv:0901.0802 [hep-ph].
%%CITATION = ARXIV:0901.0802;%%

\bibitem{Idilbi:2009cc}
A.~Idilbi, C.~Kim, and T.~Mehen,
\newblock Phys. Rev. {\bf D79}, 114016 (2009), arXiv:0903.3668 [hep-ph].
%%CITATION = ARXIV:0903.3668;%%

\bibitem{BFS}
M.~Beneke, P.~Falgari and C.~Schwinn,
\newblock (2009),
\newblock in preparation.

\bibitem{Falgari:2009zz}
P.~Falgari,
\newblock PhD thesis, RWTH Aachen University (2008),
\newblock available at
  \url{http://darwin.bth.rwth-aachen.de/opus3/volltexte/2009/2682/}.

\bibitem{Mitov:2009sv}
A.~Mitov, G.~Sterman, and I.~Sung,
\newblock Phys. Rev. {\bf D79}, 094015 (2009), arXiv:0903.3241 [hep-ph].
%%CITATION = ARXIV:0903.3241;%%

\bibitem{Becher:2009kw}
T.~Becher and M.~Neubert,
\newblock Phys. Rev. {\bf D79}, 125004 (2009), arXiv:0904.1021 [hep-ph].
%%CITATION = ARXIV:0904.1021;%%

\bibitem{Tung:1985na}
W.~K. Tung,
\newblock {\em {Group Theory in Physics}} (World Scientific, Singapore, 1985).

\bibitem{Korchemsky:1993uz}
G.~P. Korchemsky and G.~Marchesini,
\newblock Phys. Lett. {\bf B313}, 433 (1993).
%%CITATION = PHLTA,B313,433;%%

\bibitem{Beneke:1999qg}
M.~Beneke, A.~Signer and V.~A. Smirnov,
\newblock Phys. Lett. {\bf B454}, 137 (1999), [hep-ph/9903260].
%%CITATION = HEP-PH/9903260;%%

\bibitem{Kidonakis:2009ev}
N.~Kidonakis, 
  Phys.\ Rev.\ Lett.\  {\bf 102} (2009) 232003
  arXiv:0903.2561 [hep-ph].
  %%CITATION = PRLTA,102,232003;%%

\bibitem{Korchemsky:1991zp}
G.~P. Korchemsky and A.~V. Radyushkin,
\newblock Phys. Lett. {\bf B279}, 359 (1992), [hep-ph/9203222].
%%CITATION = HEP-PH/9203222;%%

\bibitem{Korchemsky:1992xv}
G.~P. Korchemsky and G.~Marchesini,
\newblock Nucl. Phys. {\bf B406}, 225 (1993), [hep-ph/9210281].
%%CITATION = HEP-PH/9210281;%%

\bibitem{Catani:1996vz}
S.~Catani and M.~H. Seymour,
\newblock Nucl. Phys. {\bf B485}, 291 (1997), [hep-ph/9605323].
%%CITATION = HEP-PH/9605323;%%

\bibitem{Becher:2009cu}
T.~Becher and M.~Neubert,
\newblock Phys. Rev. Lett. {\bf 102}, 162001 (2009), arXiv:0901.0722.
%%CITATION = ARXIV:0901.0722;%%

\bibitem{Gardi:2009qi}
E.~Gardi and L.~Magnea,
\newblock JHEP {\bf 03}, 079 (2009), arXiv:0901.1091 [hep-ph].
%%CITATION = ARXIV:0901.1091;%%

\bibitem{Becher:2009qa}
T.~Becher and M.~Neubert,
\newblock JHEP {\bf 06}, 081 (2009), arXiv:0903.1126 [hep-ph].
%%CITATION = ARXIV:0903.1126;%%

\bibitem{Moch:2004pa}
S.~Moch, J.~A.~M. Vermaseren and A.~Vogt,
\newblock Nucl. Phys. {\bf B688}, 101 (2004), [hep-ph/0403192].
%%CITATION = HEP-PH/0403192;%%

\bibitem{Beneke:1995pq}
M.~Beneke and V.~M. Braun,
\newblock Nucl. Phys. {\bf B454}, 253 (1995), [hep-ph/9506452].
%%CITATION = HEP-PH/9506452;%%

\bibitem{Beneke:1997zp}
M.~Beneke and V.~A.~Smirnov,
\newblock Nucl.\ Phys.\ {\bf B522} (1998), 321,
[hep-ph/9711391].
%%CITATION = NUPHA,B522,321;%%

\bibitem{Gatheral:1983cz}
J.~G.~M. Gatheral,
\newblock Phys. Lett. {\bf B133}, 90 (1983).
%%CITATION = PHLTA,B133,90;%%

\bibitem{Frenkel:1984pz}
J.~Frenkel and J.~C. Taylor,
\newblock Nucl. Phys. {\bf B246}, 231 (1984).
%%CITATION = NUPHA,B246,231;%%

\bibitem{Ahrens:2008nc}
V.~Ahrens, T.~Becher, M.~Neubert, and L.~L. Yang,
\newblock Eur. Phys. J. {\bf C62}, 333 (2009), arXiv:0809.4283 [hep-ph].
%%CITATION = 0809.4283;%%

\bibitem{Catani:2003zt}
S.~Catani, D.~de~Florian, M.~Grazzini and P.~Nason,
\newblock JHEP {\bf 07}, 028 (2003), [hep-ph/0306211].
%%CITATION = HEP-PH/0306211;%%

\bibitem{Idilbi:2006dg}
A.~Idilbi, X.-d. Ji and F.~Yuan,
\newblock Nucl. Phys. {\bf B753}, 42 (2006), [hep-ph/0605068].
%%CITATION = HEP-PH/0605068;%%

\bibitem{Bernreuther:2004ih}
W.~Bernreuther {\em et~al.},
\newblock Nucl. Phys. {\bf B706}, 245 (2005), [hep-ph/0406046].
%%CITATION = HEP-PH/0406046;%%

\bibitem{Mitov:2006xs}
A.~Mitov and S.~Moch,
\newblock JHEP {\bf 05}, 001 (2007), [hep-ph/0612149].
%%CITATION = HEP-PH/0612149;%%

\bibitem{MacFarlane:1968vc}
A.~J. MacFarlane, A.~Sudbery and P.~H. Weisz,
\newblock Commun. Math. Phys. {\bf 11}, 77 (1968).
%%CITATION = CMPHA,11,77;%%

\bibitem{Dokshitzer:2005ig}
Y.~L. Dokshitzer and G.~Marchesini,
\newblock JHEP {\bf 01}, 007 (2006), [hep-ph/0509078].
%%CITATION = HEP-PH/0509078;%%

\bibitem{Actis:2008rb}
S.~Actis, M.~Beneke, P.~Falgari, and C.~Schwinn,
\newblock Nucl. Phys. {\bf B807}, 1 (2009), arXiv:0807.0102 [hep-ph].
%%CITATION = ARXIV:0807.0102;%%


\bibitem{Czakon:2009zw}
M.~Czakon, A.~Mitov, and G.~Sterman,
\newblock (2009), arXiv:0907.1790 [hep-ph].
%%CITATION = ARXIV:0907.1790;%%

%\cite{Ferroglia:2009ep}
\bibitem{Ferroglia:2009ep}
A.~Ferroglia, M.~Neubert, B.~D.~Pecjak and L.~L.~Yang,
  %``Two-loop divergences of scattering amplitudes with massive partons,''
arXiv:0907.4791 [hep-ph], 
and 
%%CITATION = ARXIV:0907.4791;%%
arXiv:0908.3676 [hep-ph].
%%CITATION = ARXIV:0908.3676;%%


\end{thebibliography}
\end{document}